\pdfoutput=1

\documentclass[9pt,twocolumn]{article}
\usepackage{graphicx}
\usepackage{textcomp}
\usepackage{amsmath}
\usepackage[hmargin=1.6cm,vmargin=2.0cm]{geometry}
\usepackage[font=small,margin=0.4cm,figurename=FIGURE]{caption}
\usepackage[sort&compress,numbers]{natbib}
\usepackage{abstract}
\usepackage{hyperref}

\hypersetup{
    colorlinks=true,
    linkcolor=black,
    citecolor=black,
    filecolor=black,
    urlcolor=black,
}

\renewcommand{\abstractname}{}    
\renewenvironment{abstract}
 {\small
  \begin{center}
  \bfseries \abstractname\vspace{-1.05cm}\vspace{0pt}
  \end{center}
  \list{}{
    \setlength{\leftmargin}{1.9cm}%
    \setlength{\rightmargin}{\leftmargin}%
  }%
  \item\relax}
 {\endlist}

\begin{document}

\title{\large \bf Pathway towards programmable wave anisotropy in cellular metamaterials
}

\author{\normalsize Paolo Celli, Weiting Zhang, Stefano Gonella$^{*}$}
\date{{\small \emph{Department of Civil, Environmental, and Geo- Engineering,\\
\vspace{-3px}
University of Minnesota, Minneapolis, MN 55455, USA}}\\
\vspace{2px}
{\small $^{*}$sgonella@umn.edu}\\
\vspace{-2px}
\vspace{15px} \normalsize{{\bf \color{red} \underline{Published article}}: \emph{Phys. Rev. Appl.} {\bf 9} (1), 014014 (2018)\\\url{https://doi.org/10.1103/PhysRevApplied.9.014014}}
}

\twocolumn[
\begin{@twocolumnfalse}
\maketitle
\begin{abstract}

In this work, we provide a proof-of-concept experimental demonstration of the wave control capabilities of cellular metamaterials endowed with populations of tunable electromechanical resonators. Each independently tunable resonator comprises a piezoelectric patch and a resistor-inductor shunt, and its resonant frequency can be seamlessly re-programmed without interfering with the cellular structure's default properties. We show that, by strategically placing the resonators in the lattice domain and by deliberately activating only selected subsets of them, chosen to conform to the directional features of the beamed wave response, it is possible to \emph{override} the inherent wave anisotropy of the cellular medium. The outcome is the establishment of tunable spatial patterns of energy distillation resulting in a non-symmetric correction of the wavefields.

\vspace{0.5cm}
\end{abstract}
\end{@twocolumnfalse}
]

\subsubsection*{INTRODUCTION}
\label{sec:intro}
Cellular solids are porous media known to display unique combinations of complementary mechanical properties, such as high stiffness and high strength at low densities~\cite{Fleck_PRSA_2010}. \emph{Lattice materials}---cellular solids with ordered architectures---are obtained by spatially tessellating a fundamental building block (unit cell) comprising simple slender structural elements such as beams, plates or shells. Advances in additive manufacturing have recently propelled a resurgence of architected cellular solids as mechanical metamaterials with unprecedented functionalities at multiple scales~\cite{Schaedler_SCIENCE_2011, Zheng_SCIENCE_2014, Meza_PNAS_2015, Coulais_NATURE_2016}. Examples include fully-recoverable, energy absorbing lattices with bucklable struts~\cite{Shan_ADMA_2015, Restrepo_EML_2015}, pentamode fluid-like materials that behave as ``unfeelability'' cloaks~\cite{Buckmann_NATCOMM_2014}, lattices with negative Poisson's ratio~\cite{Lakes_ADMA_1993}, negative thermal expansion~\cite{Wang_PRL_2016}, and smart lattices with programmable stiffness~\cite{Song_ADEM_2016}.

Lattice structures also display unique dynamic properties. They commonly feature Bragg-type bandgaps as a result of their periodicity and occasionally subwavelength bandgaps for special unit cell designs or in the presence of internal resonators, thus behaving as frequency-selective stop-band filters for acoustic~\cite{Kroedel_PRAPP_2016}, elastic~\cite{Phani_JASA_2006, Baravelli_JSV_2013, Kroedel_AEM_2014, Junyi_IJSS_2016, Miniaci_APL_2016, Matlack_PNAS_2016, Warmuth_SCIREP_2017} and electromagnetic waves~\cite{Chernow_APL_2015}. They also display elastic wave anisotropy, which manifests as pronounced beaming of the energy according to highly directional patterns~\cite{Langley_JSV_1996, Ruzzene_SMS_2003, Phani_JASA_2006, Wen_JPD_2008, Carta_IJSS_2014, Celli_JSV_2014, Wang_JPD_2014, Trainiti_IJSS_2016, Zelhofer_IJSS_2017, Ganesh_APL_2017, Lefebvre_PRL_2017}. This behavior can be attributed to the fact that, at the cell scale, elastic waves are forced to propagate along the often tortuous pathways dictated by the links/struts. The spatial characteristics, symmetry landscape and frequency dependence of the anisotropic patterns are dictated by the unit cell's architecture~\cite{Celli_JAP_2014, Krattiger_AIP_2016} and are usually irreversibly determined during the design and fabrication stages. To endow cellular solids with functional flexibility and active spatial wave management capabilities, we need our structural systems to be \emph{tunable} or \emph{programmable}~\cite{Ruzzene_JVA_2000, Shan_ADFMA_2014, Nouh_JIMSS_2016, Zhu_APL_2016, Zhu_JASA_2016, Croenne_JASA_2016, Wang_ADMA_2016, Ouisse_SMS_2016, Chen_PRAPP_2017, Bilal_ADVMAT_2017}. To ensure that geometry and material requirements imposed by other functional constraints are preserved during the tuning process, it is especially important to devise minimally invasive tunability strategies~\cite{Celli_APL_2015}.

In this work, we propose and implement a strategy for tailorable spatial wave management in architected cellular solids, based on the interplay between the inherent anisotropy of the underlying lattice medium and the resonant behavior of tunable resonators strategically located on selected lattice links. Let us recall here that a resonator attached to a structural medium interacts with an incident propagating wave by distilling from the wave spectrum an interval of frequencies comprised within the neighborhood of its resonant frequency~\cite{Liu_SCIENCE_2000, Cardella_SMS_2016}. This behavior can be explained by invoking the destructive interference mechanisms between the incident wave and the wave that the resonator re-radiates, which experiences a 180$^{\mathrm{o}}$ phase shift for frequencies immediately above resonance~\cite{Lemoult_NATPHYS_2013}. As a result, by tuning the resonators as to induce distillation at frequencies for which the lattice displays anisotropic directional wavefields, and by activating selected spatial subsets of resonators located along the dominant energy beams of the directional fields, we can override selected spatial wave features. The result is a smart lattice structure whose spatial wavefields can be programmed to display several \emph{complementary} pattern corrections that \emph{relax} the symmetry of the response. For this task, we resort to resonators consisting of thin, minimally invasive piezoelectric patches shunted with resistor-inductor (RL) circuits to realize resonant RLC units, adapting and perfecting a framework previously established for beams, plates and waveguides~\cite{Hagood_JSV_1991, Casadei_SMS_2010, Airoldi_NJoP_2011, Casadei_JAP_2012, Bergamini_ADMA_2014, Wen_JIMSS_2016, Cardella_SMS_2016, Sugino_SMS_2017, Collet_IEEESensors_2014}. The resonant frequency of each electromechanical resonator can be seamlessly varied by modifying the electrical impedance of the corresponding shunting circuit, which is carried out by simply tuning one of the circuit components.

\subsubsection*{EXPERIMENTAL SETUP}
\label{sec:exp}
\begin{figure} [!htb]
\centering
\includegraphics[scale=1.45]{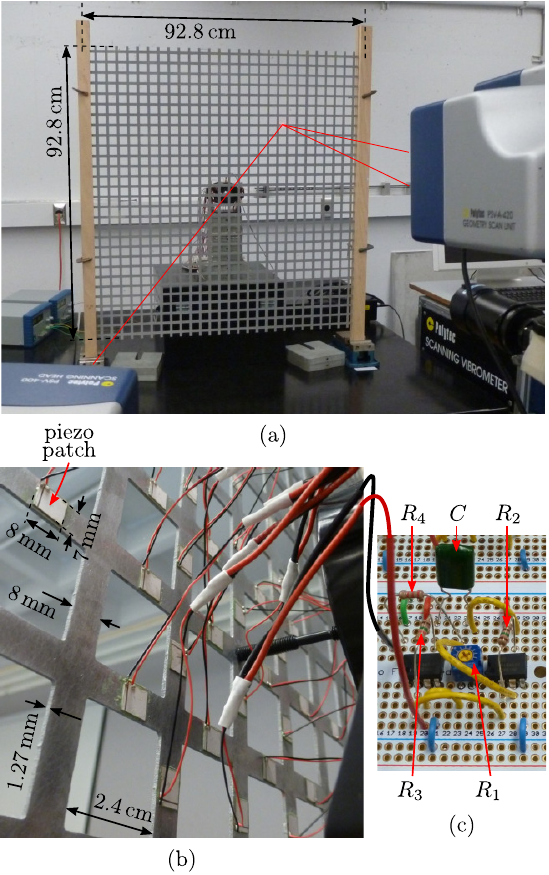}
\caption{(a) Experimental setup. (b) Detail of the rear face of the specimen near the actuation point, showing multiple piezoelectric patches bonded to the lattice structure. (c) One of the synthetic inductor circuits, schematically connected to one of the patches. Please consult the Supplemental Material~\cite{suppl} for details on the patch-circuit connection.}
\label{fig:exp}
\end{figure}

Our experimental setup is illustrated in Fig.~\ref{fig:exp}a.
The cellular medium of choice is a square lattice specimen, comprising 29$\times$29 unit cells, manufactured out of a $1.27\,\mathrm{mm}$-thick 6061 aluminum plate (Young's modulus $E=68\,\mathrm{GPa}$, density $\rho=2700\,\mathrm{kg/m^3}$, Poisson's ratio $\nu=0.33$) using water jet cutting. The characteristic length of the unit cell is $L=3.2\,\mathrm{cm}$, and the width of a lattice link is $b=8\,\mathrm{mm}$. The out-of-plane motion of points on the front face of the specimen that correspond to a pre-determined scanning grid is measured via a 3D Scanning Laser Doppler Vibrometer (3D-SLDV). The excitation signals are imparted through an electromechanical shaker and a stinger, connected to the lattice node at the center of the specimen. A detail of the rear face of the specimen near the actuation location, shown in Fig.~\ref{fig:exp}b, highlights the presence of multiple piezoelectric patches ($8\times7\times0.2\,\mathrm{mm}$ wafers made of PZT-5A) bonded to the structure. A total of 28 patches are located as to form a ring around the excitation point. All the patches are wired, but only up to seven (or eight) are simultaneously activated---i.e. connected to a RL circuit---during each experiment. Eight synthetic inductors (Antoniou circuits) are built on a solderable breadboard to serve as equivalent inductors for the eight required shunting circuits; one of them is shown in Fig.~\ref{fig:uc}c. Synthetic inductors are a staple in the shunting literature due to their compact dimensions (even for large values of inductance), versatility and tunability~\cite{Wang_SMS_2010}. DC power supplies are used to power the Op-Amps in the Antoniou circuits. To seamlessly program the synthetic inductor (and consequently the characteristics of the resonator), the resistor $R_1$ in each circuit is a tunable potentiometer; modifying $R_1$ changes the equivalent inductance $L_{eq}=R_1R_3R_4C/R_2$ and, therefore, the resonant frequency $f_r=1/[2\pi(C^pL_{eq})^{1/2}]$, where $C^p$ is the capacitance of the piezo patch. The circuit components are selected as to allow the resonators to be tuned at any frequency in the interval from $2$ to $13\,\mathrm{kHz}$. Series resistors are required to introduce enough damping to alleviate the effects of circuit instabilities~\cite{dellISOLA_SMS_2004}. More details on the experimental setup, on the circuits and on their tuning are reported in the Supplemental Material~\cite{suppl}.

\subsubsection*{NUMERICAL AND EXPERIMENTAL WAVE RESPONSE}
\label{sec:uc}
\begin{figure} [!htb]
\centering
\includegraphics[scale=1.45]{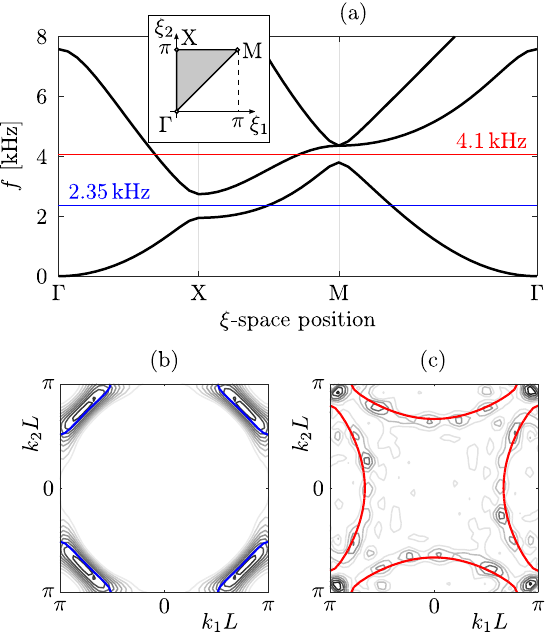}
\caption{(a) Band diagram of the pristine square lattice computed via a FE-based unit cell analysis. This diagram only comprises flexural wave modes, i.e., those relevant in the case of low-frequency, out-of-plane loads. The Irreducible Brillouin zone is shown in the insert. Colored horizontal lines highlight frequencies where wave anisotropy is most pronounced. (b) The blue lines represent the numerically computed iso-frequency contour of the first dispersion surface at $2.35\,\mathrm{kHz}$; the underlying contours are the spectral lines computed from the experimental lattice response to a burst with carrier frequency $2.35\,\mathrm{kHz}$. (c) Same as (b) for the second dispersion surface at $4.1\,\mathrm{kHz}$.}
\label{fig:uc}
\end{figure}

The default wave response of the pristine lattice is illustrated in Fig.~\ref{fig:uc}.
The band diagram of an infinite lattice having the same cell dimensions, geometry and material properties as our specimen is shown in Fig.~\ref{fig:uc}a; this result is obtained through a Bloch analysis of a unit cell modeled with plate finite elements (implementing Mindlin's plate model) limited to wave vectors sampled along the contour of the Irreducible Brillouin Zone (IBZ)~\cite{Phani_JASA_2006}. We are especially interested in frequencies where the appearance of partial bandgaps suggests pronounced wave anisotropy. In the low-frequency range shown in Fig.~\ref{fig:uc}a, these frequencies are approximately $2.35\,\mathrm{kHz}$ and $4.1\,\mathrm{kHz}$. The blue lines in Fig.~\ref{fig:uc}b mark the Cartesian iso-frequency contour obtained by slicing the first dispersion surface at $2.35\,\mathrm{kHz}$, and highlight how, at this frequency, waves are predominantly allowed to propagate along directions that are $\pm 45^{\mathrm{o}}$-oriented with respect to the horizontal axis. Iso-frequency contours provide information analogous to the slowness curves~\cite{Tallarico_JMPS_2017}, whereby proximity to the origin implies high phase velocity, and vice versa. The iso-frequency contour of the second surface at $4.1\,\mathrm{kHz}$ displays complementary features: wave speeds are now much higher along the horizontal and vertical directions.

\begin{figure} [!htb]
\centering
\includegraphics[scale=1.45]{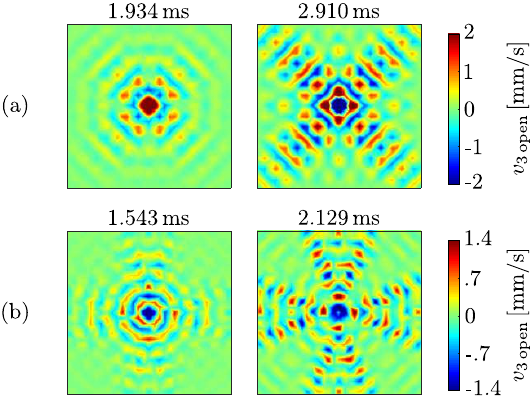}
\caption{Transient experimental response of the lattice to 13-cycle bursts with carrier frequencies (a) $2.35\,\mathrm{kHz}$ and (b) $4.1\,\mathrm{kHz}$, when all piezoelectric patches are in their open circuit configuration. }
\label{fig:open}
\end{figure}

In Fig.~\ref{fig:open}, we report the out-of-plane velocity snapshots (at two time instants) of the measured transient response of our lattice specimen. The excitation signals are 13-cycle bursts with carrier frequencies corresponding to $2.35\,\mathrm{kHz}$ and $4.1\,\mathrm{kHz}$, respectively. Note that, at this stage, all patches are in their open circuit configuration.
The response at $2.35\,\mathrm{kHz}$ (Fig.~\ref{fig:open}a) shows that, at this frequency, the wavefields feature four highly-beamed packets propagating along $\pm 45^{\mathrm{o}}$-oriented directions, as predicted by the iso-frequency contour in Fig.~\ref{fig:uc}b. On the other hand, the wavefields corresponding to $4.1\,\mathrm{kHz}$ (Fig.~\ref{fig:open}b) feature packets propagating mainly along the vertical and horizontal directions---consistent with the predictions of Fig.~\ref{fig:uc}c. In Figs.~\ref{fig:uc}b-c, to further demonstrate the excellent agreement between the results from the unit cell analysis and the laser-acquired experimental data, we superimpose to the iso-frequency contours the spectral lines of the lattice response at the same frequencies, obtained from the experimentally acquired wavefields through a 2D Discrete Fourier Transform procedure (2D-DFT; see the SM section for details on the reconstruction procedure). This comparison also highlights how the presence of the open circuit (non-shunted) patches has minimal influence on the characteristics of the medium in this low-frequency regime.

\subsubsection*{WAVE-RESONATOR INTERACTION}
\label{sec:res}
To elucidate the mechanisms behind the wave-resonator interaction in the specific case where the resonator comprises a piezoelectric patch shunted with a RL circuit, we resort to a simple one dimensional experiment. The specimen is a $117.4\,\mathrm{cm}$-long and $8\,\mathrm{mm}$-wide beam with the same thickness and material properties as the lattice specimen. The beam is clamped at both ends, and the excitation is imparted near one of the clamps. Approximately at the center of the beam, we bond a single piezoelectric patch (identical to those used in the lattice experiment). A schematic of the beam, with illustration of the location where the response is measured by the 3D-SLDV system, is shown in Fig.~\ref{fig:res}a (the setup for this experiment is shown in Fig.~S7 and discussed in the Supplemental Material~\cite{suppl}).
\begin{figure} [!htb]
\centering
\includegraphics[scale=1.45]{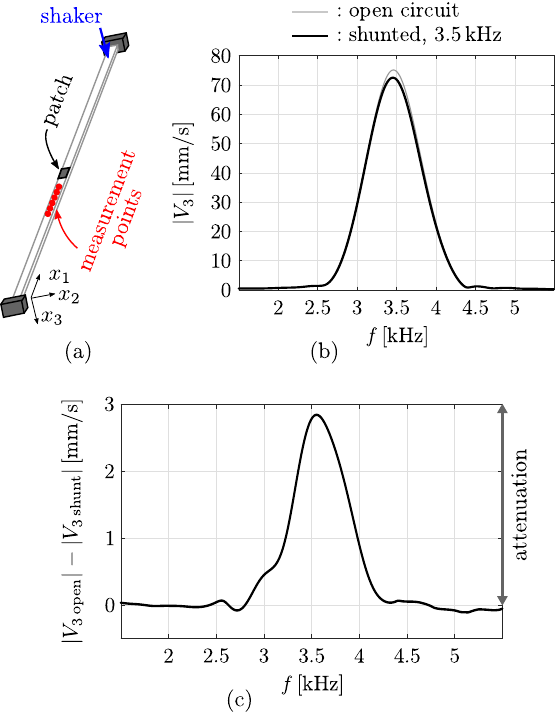}
\caption{Experimental results on wave-resonator interaction. (a) Schematic of the experimental setup (clamped-clamped beam with a single piezoelectric patch). (b) Frequency spectra of the filtered out-of-plane velocity time histories for open-circuit and shunted cases, in response to a burst with $3.5\,\mathrm{kHz}$ carrier frequency. The curves are obtained by averaging the spectra recorded at multiple measurement points downstream from the patch. (c) Difference between the open circuit spectrum and the shunted one.}
\label{fig:res}
\end{figure}
The beam is excited via a 9-cycle burst with carrier frequency $3.5\,\mathrm{kHz}$. Boundary reflections are eliminated through a time-filtering procedure, also discussed in~\cite{suppl}. In Fig.~\ref{fig:res}b, we compare the frequency spectra of the out-of-plane velocity signals recorded in the open circuit case (gray line) and in the shunted case (black line). These signals have been obtained by averaging the response at multiple measurement points located after the resonator. The patch-circuit system is tuned as to resonate around $3.5\,\mathrm{kHz}$, although a small error in pinpointing the frequency is often expected. We observe that, as predicted, the resonator causes wave attenuation/distillation in the neighborhood of the tuning frequency. The amplitude of attenuation is small compared to other instances reported in the literature on shunted systems due to the fact that here the wave packet is purely incident and interacts with the resonator only once. Stronger attenuation results usually arise from multiple wave-resonator interactions, as in steady-state conditions or when we can aggregate the effects of multiple transient packets bouncing back and forth between the structure's boundaries~\cite{Cardella_SMS_2016}. The action of the resonator can be better visualized by subtracting the shunted spectrum from the open circuit one; this differential plot is shown in Fig.~\ref{fig:res}c. Notably, most of the attenuation is recorded right above the expected resonance. This behavior can be qualitatively explained using arguments of phase delay and wave interference. Specifically, when a propagating wave interacts with a resonator, a fraction of the energy is stored in the resonator and re-radiated into the structure, possibly with a phase shift. This wave can interact constructively or destructively with the incident wave, according to their relative phase~\cite{Liu_SCIENCE_2000, Lemoult_NATPHYS_2013, JIN_PRB_2017}; above resonance, where incident and re-radiated wave are in opposition of phase, destructive interference mechanisms result in signal attenuation. This effect becomes less pronounced as we move away from resonance, the resonator becomes progressively less engaged and the portion of the reradiated energy drops, ultimately dictating the width of the bandgap. Note that, while this effect is precisely observed for continuous harmonic excitation (which explains why the onset of resonant bandgaps in the frequency domain can be pinpointed experimentally with great accuracy in steady state conditions), its signature is blurrier for burst excitations with compact support where issues of packet delay and distortion may contaminate the inference.


\subsubsection*{ANISOTROPY OVERRIDING}
\label{sec:anis}
\begin{figure*} [!htb]
\centering
\includegraphics[scale=1.45]{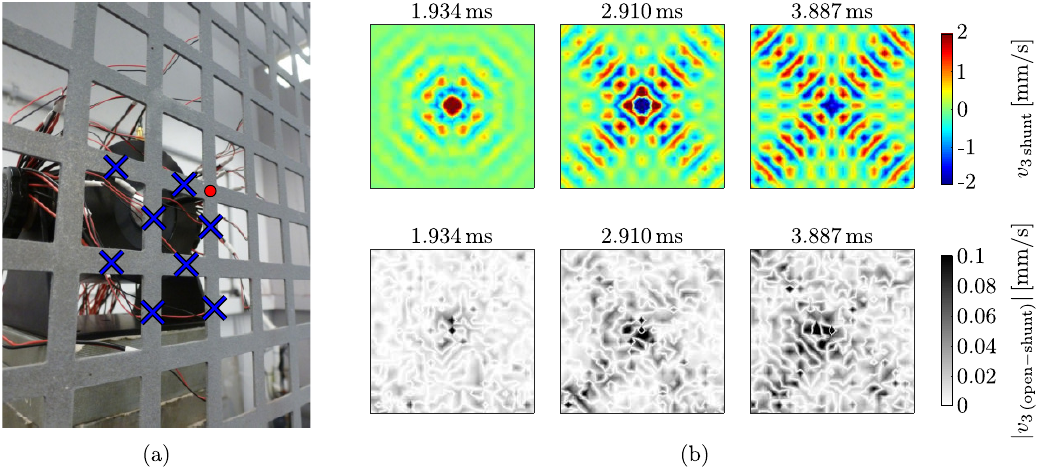}
\caption{Anisotropy overriding at $2.35\,\mathrm{kHz}$; the targeted wave feature is the packet propagating towards the bottom-left corner of the specimen. (a) Detail of the front (scanned) face of the specimen; the circular marker indicates the actuation location and the crosses denote the links with shunted piezos. (b) Top row: wavefields acquired at three time instants. Bottom row: spatial patterns displayed by the difference between the open circuit and the shunted response.}
\label{fig:resBR}
\end{figure*}

\begin{figure} [!htb]
\centering
\includegraphics[scale=1.45]{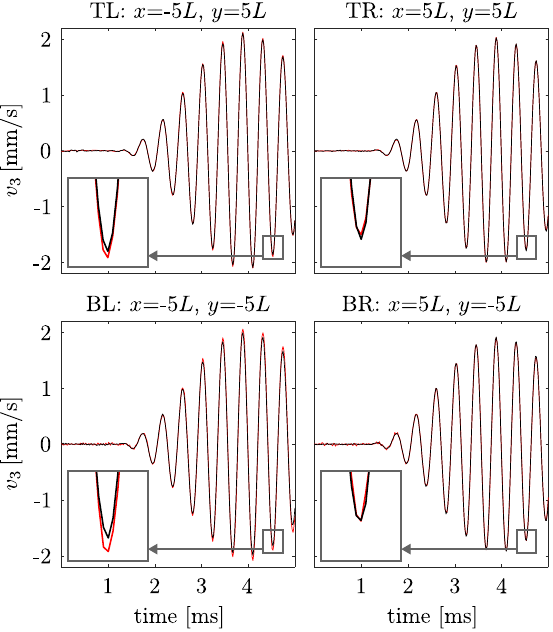}
\caption{Clockwise: time histories of the open circuit (red line) and shunted (black line) response at four points in the top-left (TL), top-right (TR), bottom-left (BL) and bottom-right (BR) quadrants, respectively. The targeted wave feature is the packet propagating towards the bottom-left corner of the specimen.}
\label{fig:resBRth}
\end{figure}

At this stage, we know how the lattice responds to bursts at several frequencies of interest and how a traveling wave interacts with an electromechanical resonator based on a shunted piezo patch. Therefore, we have all the ingredients to investigate the influence of strategically placed, properly tuned resonators on specific wave features of the anisotropic response of the lattice. A numerical demonstration of this strategy for an idealized lattice of springs and masses with mass-in-mass nodes, which can be seen as a purely mechanical analog of our system, is reported in the Supplemental Material~\cite{suppl}. Despite the pronounced modeling differences, these simulations are insightful in that they allow to freely explore the parameter space of the population of resonators far beyond the constraints of the experiments. Back to our experiments, we first consider the lattice response at $2.35\,\mathrm{kHz}$, shown in Fig.~\ref{fig:open}a. Our goal is to override one of the four $\pm 45^{\mathrm{o}}$-oriented packets propagating from the excitation point. In our first test, we shunt eight patches located along the path of the packet propagating towards the bottom-left corner of the domain. The selected patches, located on the links highlighted by cross-shaped markers in Fig.~\ref{fig:resBR}a, are connected to the synthetic inductors on the circuit board through series resistors with resistance $R_s=1\,\mathrm{k\Omega}$.
Each resonator is tuned at $2.2\,\mathrm{kHz}$ (see the Supplemental Material for details on the tuning procedure~\cite{suppl}); we choose this lower tuning frequency rather than the nominal $2.35\,\mathrm{kHz}$ to compensate for the shift in resonance caused by the large value of $R_s$ (necessary to avoid circuit instabilities). The response of the lattice at three distinct time instants is shown in the top row of Fig.~\ref{fig:resBR}b. At first glance, we do not detect any macroscopic modification with respect to the open circuit response of Fig.~\ref{fig:open}a. To further explore the data and reveal potential higher-order correction effects in the shunted response, in the bottom row of Fig.~\ref{fig:resBR}b, we report the differential wavefields obtained by subtracting the shunted response from the open circuit one at the considered time instants. Upon this operation, an asymmetric pattern consistent with our expectations unfolds. As predicted, properly-tuning a selected subset of resonators produces a frequency-distillation (and consequently an amplitude correction) of the wave packet in the spatial neighborhood of the activated resonators: packets traveling from the source towards the bottom-left corner of the domain carry the strongest signature of shunting. In Fig.~\ref{fig:resBRth}, we report the velocity time histories recorded at four characteristic locations on the specimen's surface---one point in each quadrant of the scanned area (top-right, bottom-right, bottom-left and top-left).
These results confirm that waves propagating towards the bottom-left corner are the most affected by the resonators and indeed experience wave attenuation. These and other results in this Section, albeit representing an unequivocal proof of concept of the anisotropy overriding capabilities of shunted lattices, reflect effects that are one order of magnitude smaller than the amplitude of the wave response. Stronger attenuation results could ostensibly be attainable through refinements of the experimental setup, e.g., by using a larger number of simultaneously activated resonators or by working with improved circuitry with reduced parasitic resistance. These technological improvements will be addressed in future investigations.

\begin{figure*} [!htb]
\centering
\includegraphics[scale=1.45]{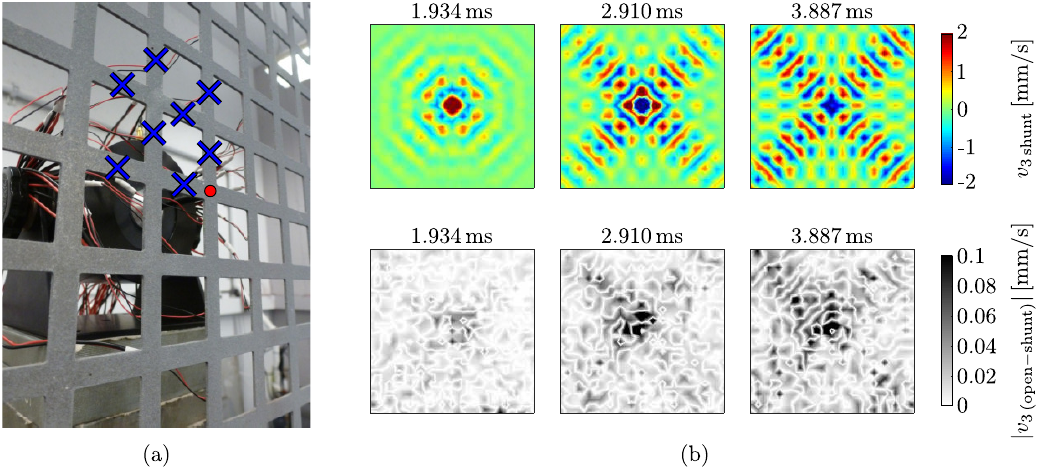}
\caption{Anisotropy overriding at $2.35\,\mathrm{kHz}$; the targeted wave feature is the packet propagating towards the top-left corner of the specimen. (a) Detail of the front (scanned) face of the specimen, indicating where the shunted patches are located. (b) Top row: wavefields acquired at three time instants. Bottom row: spatial patterns displayed by the difference between the open circuit and the shunted response.}
\label{fig:resTR}
\end{figure*}

To demonstrate the tunability of our metamaterial system, we re-program the lattice by shunting a different set of patches and re-tuning the circuits as to override the wave packet propagating, still at $2.35\,\mathrm{kHz}$, towards the top-left corner of the domain. Note that re-tuning is needed since each patch has a slightly different capacitance. This experiment is illustrated in Fig.~\ref{fig:resTR}.
In this second scenario, we again manage to override the desired feature of the anisotropic wave pattern.

\begin{figure*} [!htb]
\centering
\includegraphics[scale=1.45]{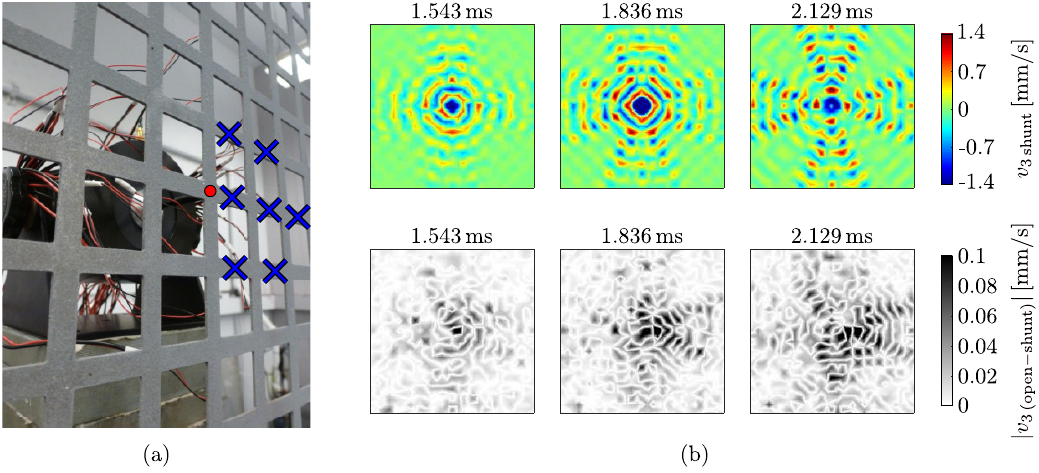}
\caption{Anisotropy overriding at $4.1\,\mathrm{kHz}$; the targeted wave feature is the rightward-propagating packet. (a) Detail of the front (scanned) face of the specimen, indicating where the shunted patches are located. (b) Top row: wavefields acquired at three time instants. Bottom row: spatial patterns displayed by the difference between the open circuit and the shunted response.}
\label{fig:resL}
\end{figure*}

We now shift our attention towards the wave response at $4.1\,\mathrm{kHz}$---characterized by complementary wave patterns propagating along the vertical and horizontal directions as illustrated in Fig.~\ref{fig:open}b. By shunting seven patches located on the links immediately to the right of the excitation point as indicated in Fig.~\ref{fig:resL}a, we program the system to override the right-going wave packet. Note that, in this case, the resonators are tuned exactly at $4.1\,\mathrm{kHz}$, since the required $330\,\mathrm{\Omega}$ series resistance seems not to produce any appreciable shift in resonance frequency.
The differential wavefields in Fig.~\ref{fig:resL}b show that the right-going horizontal feature is indeed the one being attenuated by the resonators. An additional result targeting another direction of propagation is reported for completeness in the Supplemental Material.

\subsubsection*{CONCLUSIONS}
\label{sec:concl}
In this work, we have demonstrated that the frequency-selective anisotropic wave patterns intrinsically established in cellular metamaterials can be corrected by resorting to clouds of strategically placed, tunable electromechanical resonators and by selecting objective-specific spatial activation strategies. We have shown that we can achieve levels of programmability that allow to override the anisotropy of the wave response along different directions and at different frequencies. It is reasonable to assume that results similar to those shown in this proof-of-concept study for a simple square lattice would be achievable in more complex two- and three-dimensional architected cellular solids, albeit requiring even more complex electromechanical control systems. Integrated manufacturing of the electromechanical resonators, improved circuitry~\cite{FloresParra_SMS_2017}, or the adoption of less invasive resonators capable of offering a stronger coupling while simultaneously allowing the activation of more than seven or eight resonators could increase the tangibility of the observed effects. This kind of developments will eventually pave the way towards families of smart lattices with programmable spatial wave control capabilities such as wave steering, energy channeling and re-routing, and spatial filtering.

\subsubsection*{ACKNOWLEDGEMENTS}
We acknowledge the support of the National Science Foundation (grant CMMI-1266089). PC acknowledges the support of the University of Minnesota through the Doctoral Dissertation Fellowship. We also thank Davide Cardella for his contribution during the initial stages of the project.

\begin{center}
\noindent\rule{3cm}{0.4pt}
\end{center}

\renewcommand\refname{\vskip -1.3cm}
\bibliographystyle{unsrt}

\clearpage

\setcounter{figure}{0}
\setcounter{page}{1}
\renewcommand{\thefigure}{S\arabic{figure}}
\renewcommand{\theequation}{S\arabic{equation}}
\renewcommand{\thepage}{S\arabic{page}}

\onecolumn
\section*{\Large Supplemental material (SM)}

\subsubsection*{ADDITIONAL DETAILS ON THE LATTICE EXPERIMENT}

\begin{figure} [!htb]
\centering
\includegraphics[scale=1.45]{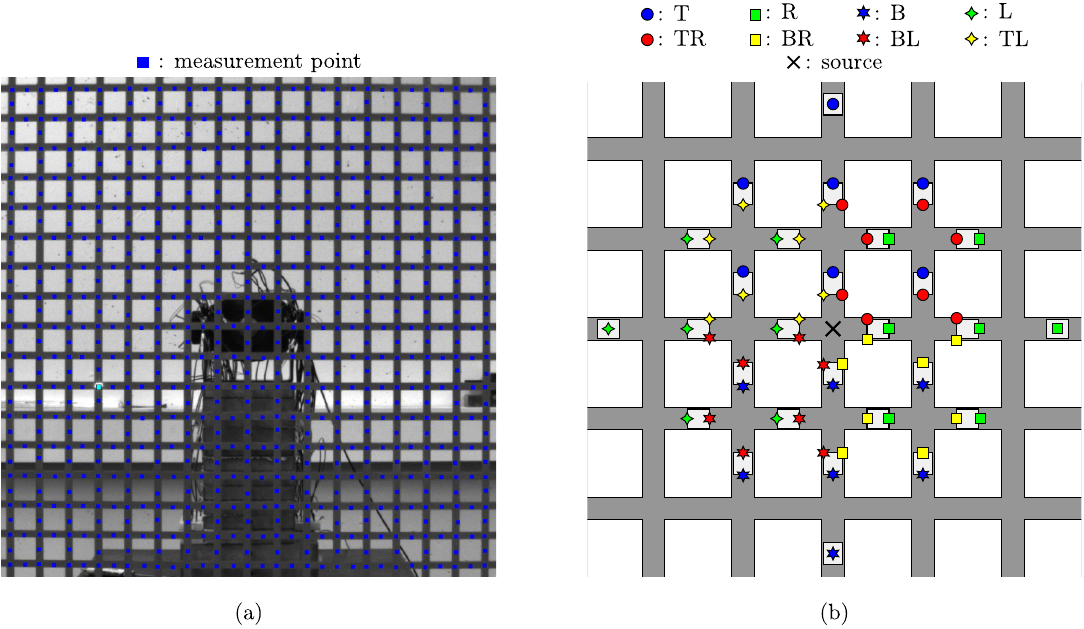}
\caption{(a) Picture of the specimen taken from the 3D-SLDV camera. The blue square markers denote the measurement locations. (b) Sketch of a portion of the rear face of the specimen, near the excitation location. The light colored rectangles represent piezoelectric patches. Patches are labeled according to the subsets they belong to, which reflect the different directional activation strategies.}
\label{fig:grid}
\end{figure}

In this Section, we report additional details on the experimental setup used to measure the response of the square lattice. Both signal generation and acquisition are performed through a 3D Scanning Laser Doppler Vibrometer system (3D-SLDV, Polytec PSV-400-3D). The 13-cycle burst signals are generated in MATLAB by windowing (Hann window) a sinusoidal signal, and by appending a ``relaxation time'' (the relaxation time, whose length is 150 times the burst length, is needed to allow the lattice structure to attain an undeformed state between consecutive measurements). The frequency of the sinusoid is the carrier frequency of the burst. The signals generated by the SLDV system have an amplitude of $400\,\mathrm{mV}$ and are amplified through a Br\"uel \& Kj\ae r Type $2718$ Power Amplifier, whose gain is set to $20\,\mathrm{dB}$. The amplified signal is then transmitted to an electromechanical shaker (Br\"uel \& Kj\ae r Type $4810$), that uses a stinger to impart the excitation to the center node of the specimen. In-plane and out-of-plane velocities are measured at all the locations highlighted by blue square markers in Fig.~\ref{fig:grid}a. 
Due to the excitation configuration, we are only interested in the out-of-plane motion of the lattice. As for the acquisition procedure, measurements are repeated multiple times (5 times when the carrier frequency of the excitation signal is at $2.35\,\mathrm{kHz}$ and 10 times at $4.1\,\mathrm{kHz}$) and averaged at each measurement location to eliminate noisy features from the response. As for the vibrometer channel, we select a $5\,\mathrm{V}$ range and DC coupling. We select a sampling frequency $f_s=51.2\,\mathrm{kHz}$ and $4096$ sampling points; as a result, the time step is $\Delta t=19.531\,\mathrm{\mu s}$ and the total acquisition time is $80\,\mathrm{ms}$. An internal trigger is used to make sure there is no phase lag between the time histories recorded at different scanning points.

As we already mentioned in the main article, the square lattice is cut from a $1.27\,\mathrm{mm}$-thick 6061 aluminum plate via water jet cutting. The material properties of 6061 Al are: Young's modulus $E=68\,\mathrm{GPa}$, density $\rho=2700\,\mathrm{kg/m^3}$, Poisson's ratio $\nu=0.33$. The lattice comprises $29\times 29$ unit cells with characteristic dimension $L=3.2\,\mathrm{mm}$; the width of a lattice link is $8\,\mathrm{mm}$. A total of 28 piezoelectric patches (STEMiNC, part number SMPL7W8T02412WL) are bonded to the lattice links in the neighborhood of the excitation location, as shown in Fig.~\ref{fig:grid}b (the patches are the light gray rectangles). In this schematic, patches are labeled according to the subset(s) to which they belong. Sets T, R, B, L are used to target the upward-, rightward-, downward-, leftward- propagating wave packets of the anisotropic wavefield at $4.1\,\mathrm{kHz}$, respectively. Sets TR, BR, BL, TL are used to target the packets of the anisotropic wavefield at $2.35\,\mathrm{kHz}$ propagating towards the top-right, bottom-right, bottom-left and top-left corners of the lattice, respectively. The patches are carefully bonded to the lattice with a 2-part epoxy glue (3M Scotch-Weld 1838 B/A).

\begin{figure} [!htb]
\centering
\includegraphics[scale=1.45]{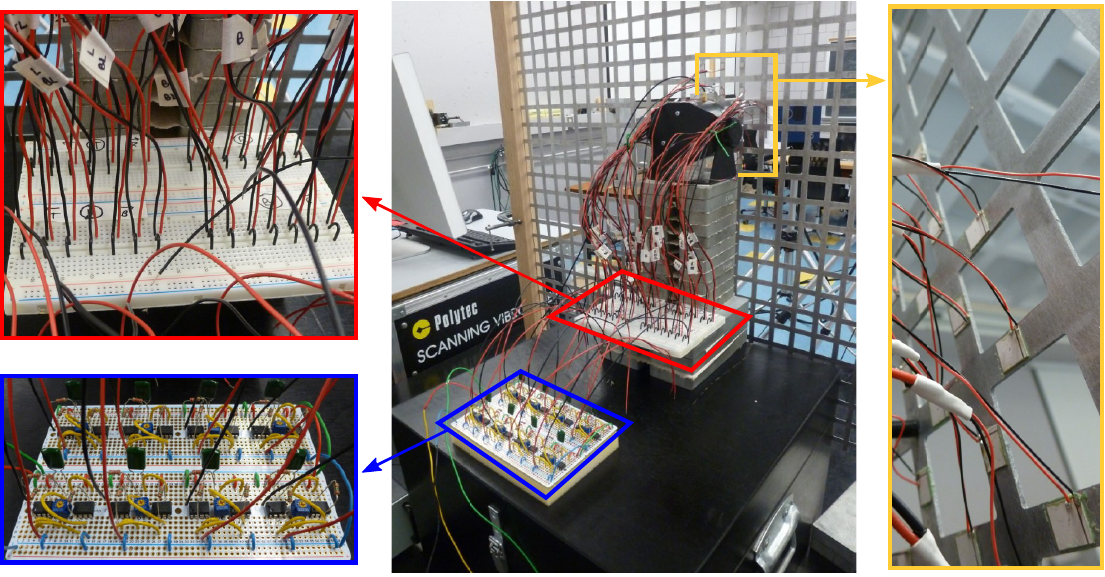}
\caption{Rear face of the specimen. The details illustrate the patches and how they are wired (yellow box), the intermediate breadboard where wires coming from the circuits and patches are connected through series resistors (red box), and the circuit board with the synthetic inductors (blue box).}
\label{fig:wires}
\end{figure}

Two wires, one per electrode, stem from each piezoelectric patch as shown in the yellow-boxed detail of Fig.~\ref{fig:wires}. This multitude of wires is connected to an intermediate breadboard (boxed in red),
where the wires coming from the synthetic inductors (built on solderable breadboards as shown in the blue box) and from the patches are connected.

\subsubsection*{ADDITIONAL DETAILS ON THE CIRCUITRY}

\begin{figure} [!htb]
\centering
\includegraphics[scale=1.45]{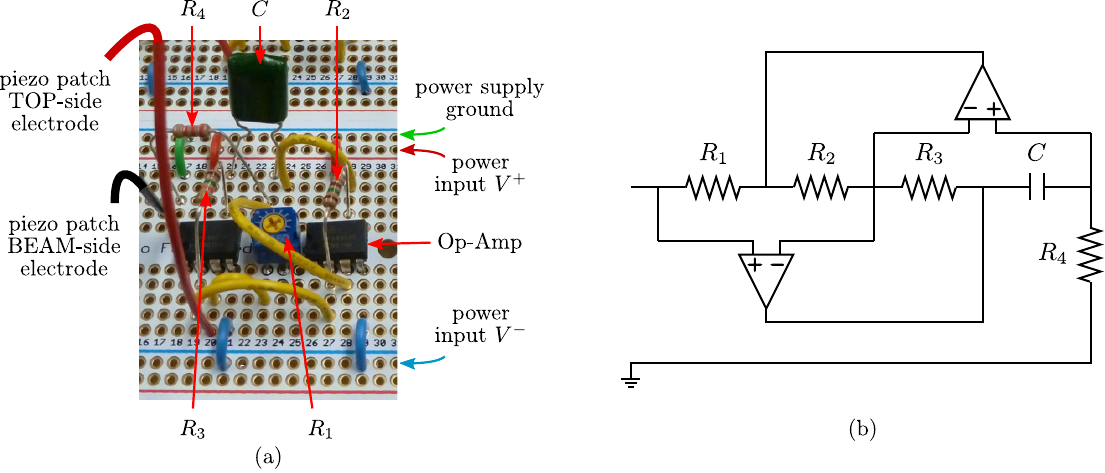}
\caption{(a) Antoniou circuit built on a solderable breadboard. (b) Circuit schematic.}
\label{fig:circ}
\end{figure}

One of the Antoniou circuits, built on a solderable breadboard (Adafruit ADA571), is shown in Fig.~\ref{fig:circ}a and schematically depicted in Fig.~\ref{fig:circ}b.
Solderable breadboards are chosen due to their improved reliability with respect to conventional ones. Each circuit comprises two operational amplifiers (TI OPA445AP), powered by two DC power supplies (BK Precision 1667) connected in series as to generate a $\pm 30\,\mathrm{V}$ voltage; it also comprises three conventional resistors ($R_2$, $R_3$, $R_4$), a potentiometer ($R_1$, Copal Electronics CT6EP102-ND) and a Mylar-type capacitor ($C$). The parameters we choose are $C=0.1\,\mathrm{\mu F}$, $R_2=150\,\mathrm{\Omega}$, $R_3=1500\,\mathrm{\Omega}$, $R_4=2200\,\mathrm{\Omega}$, while $R_1$ can be varied between $25\,\mathrm{\Omega}$ and $1000\,\mathrm{\Omega}$. The equivalent inductance of the Antoniou circuit is given by
\begin{equation}
L_{eq}=\frac{R_1\,R_3\,R_4\,C}{R_2}\,\,.
\end{equation}
Given our choice of circuit parameters, $L_{eq}$ can range between $0.044\,\mathrm{H}$ and $2.2\,\mathrm{H}$. Once an Antoniou circuit is connected to a piezoelectric patch with capacitance $C^p$, the resonant frequency of the patch-circuit system can be calculated as
\begin{equation}
f_{r}^p=\frac{1}{2 \pi}\sqrt{\frac{1}{L_{eq}C^p}}\,\,.
\label{eq:f}
\end{equation} 
The 28 piezoelectric patches bonded to our lattice specimen do not all have the same capacitance; their average capacitance is $C^p_{ave}=2.76\,\mathrm{nF}$. Using this average value and the formula in Eq.~\ref{eq:f}, we can estimate that each electromechanical resonator can be tuned between $2\,\mathrm{kHz}$ and $13\,\mathrm{kHz}$.

\begin{figure} [!htb]
\centering
\includegraphics[scale=1.45]{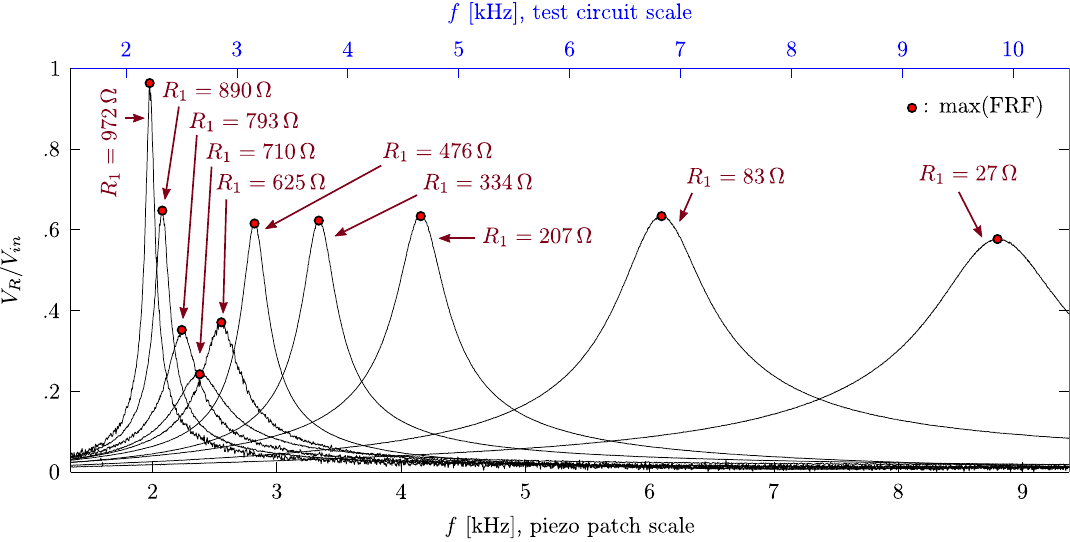}
\caption{Electrical characterization of a synthetic inductor. Each line corresponds to the circuit response for a specific value of $R_1$ (indicated in the figure). The peak of each frequency response function is highlighted by a red circular marker. Two frequency axes are reported. The test circuit scale is the original frequency scale; the piezo patch scale is derived from the test circuit one by knowing the ratio between the test capacitance and the average capacitance of the piezo patches.}
\label{fig:elec}
\end{figure}

To make sure that each one of the eight Antoniou circuits works properly, we test them electrically one by one before using them to shunt the piezo patches. This test is performed by connecting a synthetic inductor in series with a test resistor ($R^t=1000\,\mathrm{\Omega}$) and a test capacitor ($C^t=2.2\,\mathrm{nF}$, value chosen in the neighborhood of the piezo capacitance). The frequency response function of this resonant circuit, obtained in response to a multitone signal (exciting all frequencies in the $0$--$10.5\,\mathrm{kHz}$ range) as the transfer function between the input voltage ($V_{in}$) and the voltage across the resistor ($V_R$), is shown in Fig.~\ref{fig:elec}.
Each curve in this plot represents the response of the circuit for a specific value of $R_1$. Increasing $R_1$ causes the resonant frequency of the circuit to decrease. 
This plot is characterized by two frequency scales. The original scale obtained from the experimental data is the ``test circuit scale''. The other one, the ``piezo patch'' scale, is obtained by multiplying the frequency values of the test circuit scale by the correcting factor $\sqrt{C^t/C^p_{ave}}$. The latter scale gives us an idea regarding the resonance frequency of the patch-circuit system for different $R_1$ values. Fig.~\ref{fig:elec} shows that for $R_1 < 625\,\mathrm{\Omega}$ the amplitude of the resonance peak is constant and approximately equal to 0.6. For $R_1 \geq 625\,\mathrm{\Omega}$, on the other hand, the amplitude of the resonance peak is not constant; moreover, the lines corresponding to these values of $R_1$ are considerably more noisy, thus indicating a less stable behavior. In light of these considerations and by looking at the piezo patch scale, we can understand that resonators tuned at $2.35\,\mathrm{kHz}$ (the frequency corresponding to one of the two anisotropic wave patterns of interest) will be less effective than resonators tuned at $4.1\,\mathrm{kHz}$ (corresponding to the other anisotropic wave pattern of interest).

\begin{figure} [!htb]
\centering
\includegraphics[scale=1.45]{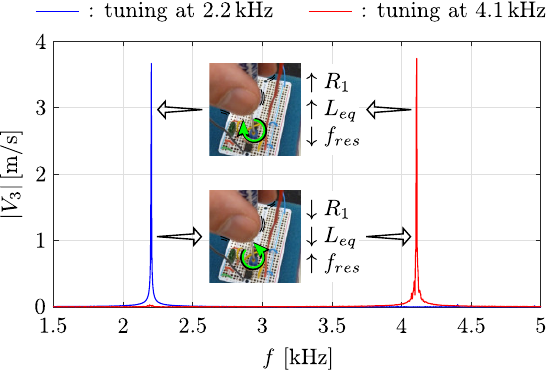}
\caption{Illustration of the phenomenon of circuit instabilities, manifesting as self-excited single frequency vibrations at the resonance of the patch-circuit system. The two lines correspond to different tuning frequencies. The inserts show how to adjust the potentiometer in order to increase or decrease the resonant frequency.}
\label{fig:tuning}
\end{figure}

To tune the circuits, and to make sure that each shunt-patch system is properly tuned at the desired frequency, we are typically required to test each synthetic inductor electrically (using the same test circuit discussed above) every time we want to re-program our system. The main drawback of this electrical tuning procedure is that it heavily relies on the knowledge of the precise values of the patch capacitance [F. A. C. Viana and V. Steffen Jr, \emph{J. Braz. Soc. Mech. Sci. Eng.} {\bf 28}, 293 (2006)]. Moreover, since all patches have a different capacitance, it is extremely inconvenient to re-test a circuit electrically every time we want to use it to shunt a different patch. In this work, we bypass this inconvenience by exploiting the electrical instabilities that arise during the shunting process to precisely tune the electromechanical resonators. When connecting a synthetic inductor to a piezoelectric patch in the absence of a series resistor $R_s$, the system is unstable, and this instability manifests as a self-excited vibration of the patch at the resonant frequency of the patch-circuit system. An audible manifestation of this instability is ``screaming''---a single-pitch noise coming from the patch. Instabilities should be avoided because they are usually associated with a temperature increase that could damage the circuit components. This behavior, seldom discussed in the literature, has also been documented by dell'Isola et al. [F. dell’Isola, C. Maurini and M. Porfiri, \emph{Smart Mater. Struct.} {\bf 13}, 299 (2004)]. For example, the blue line in Fig.~\ref{fig:tuning} represents the frequency spectrum of the signal recorded at a point of the beam when the only excitation causing it to vibrate is the self-driven instability-based vibration of the patch (no external excitation is provided). Note that the power necessary to generate these vibrations is provided by the DC power supplies that power the operational amplifiers in the synthetic inductors.
Since the peak in this frequency response corresponds precisely to the frequency at which the patch-shunt system is tuned, we can accurately tune the circuit at the desired frequency by acting on the potentiometer and by monitoring the position of this instability peak. The red line corresponds to a tuning frequency of $4.1\,\mathrm{kHz}$. If we want to change the tuning frequency from $4.1\,\mathrm{kHz}$ to $2.2\,\mathrm{kHz}$, we increase the value of $R_1$ (rotating the potentiometer clockwise) and, therefore, we increase $L_{\mathrm{eq}}$ while decreasing the resonant frequency of system. This way, we can move the instability peak towards lower frequencies until it is positioned at the desired value, as shown by the blue line in Fig.~\ref{fig:tuning}. Note that, to avoid the afore-mentioned overheating of the circuit components, it is suggested not to keep shunt and circuit connected without $R_s$ for more than few seconds.

When using the electromechanical resonators to control elastic waves, we need to use a large enough series resistance in the shunting circuit in order to avoid instabilities. The value of $R_s$ we use in the experiments at $2.35\,\mathrm{kHz}$ is $1000\,\mathrm{\Omega}$, while the one for the experiments at $4.1\,\mathrm{kHz}$ is $330\,\mathrm{\Omega}$. When we target waves with carrier $2.35\,\mathrm{kHz}$, we tune the circuits as to resonate at $2.2\,\mathrm{kHz}$. This is done to counteract the fact that series resistors of $1000\,\mathrm{\Omega}$ cause a shift of the resonance peak towards higher frequencies (while no noticeable shift was noticed when tuning at $4.1\,\mathrm{kHz}$).

\subsubsection*{ISO-FREQUENCY CONTOURS RECONSTRUCTION}
\begin{figure*} [!htb]
\centering
\includegraphics[scale=1.45]{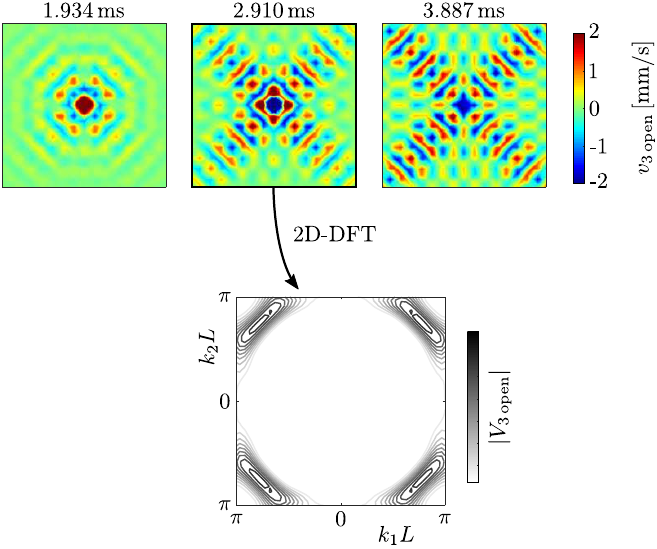}
\caption{Iso-frequency contour reconstruction at $2.35\,\mathrm{kHz}$. The three wavefield correspond to the response of the lattice, when all patches are in their open circuit configuration, to a 13-cycle burst with carrier frequency $2.35\,\mathrm{kHz}$. The contour plot at the bottom of the figure represents the experimentally-reconstructed slowness curve at $2.35\,\mathrm{kHz}$.}
\label{fig:rec}
\end{figure*}

In Section~3 of the main article, we compared theoretical and experimentally-reconstructed iso-frequency contours, or slowness curves. The reconstruction procedure is illustrated in Fig.~\ref{fig:rec}.
First of all, we excite the lattice structure (with all patches in their open circuit configuration) with a 13-cycle burst signal with carrier frequency $2.35\,\mathrm{kHz}$. Three time instants of the out-of-plane lattice response (velocity $v_3$) are shown as wavefields in Fig.~\ref{fig:rec}. We then select a time instant where the wavefield is fully developed while displays minimal boundary reflections ($2.910\,\mathrm{ms}$). Time instant $1.934\,\mathrm{ms}$, for example, is not suitable due to the fact that the wavefield is not yet well developed. On the other hand, $3.887\,\mathrm{ms}$ is not suitable either due to the fact that, at this time instant, boundary reflections have started to kick in. The 2D array representing the selected wavefield data is then analyzed through the spectrum of a 2D Discrete Fourier Transform. The result of this procedure is shown at the bottom of Fig.~\ref{fig:rec}; in this plot, the contours correspond to discrete values of the spectral amplitude $|V_3|$. Since the burst excites a band of frequencies, it is impossible to reconstruct a single iso-frequency contour line. However, due to the fact that the burst is designed to give higher spectral amplitude to the carrier frequency, $2.35\,\mathrm{kHz}$ in this case, it is safe to assume that the experimentally-reconstructed slowness curve should go through the highest-amplitude $|V_3|$ contours. The same procedure is repeated for $4.1\,\mathrm{kHz}$.

\subsubsection*{ADDITIONAL DETAILS ON THE WAVE-RESONATOR INTERACTION EXPERIMENTS}

\begin{figure*} [!htb]
\centering
\includegraphics[scale=1.45]{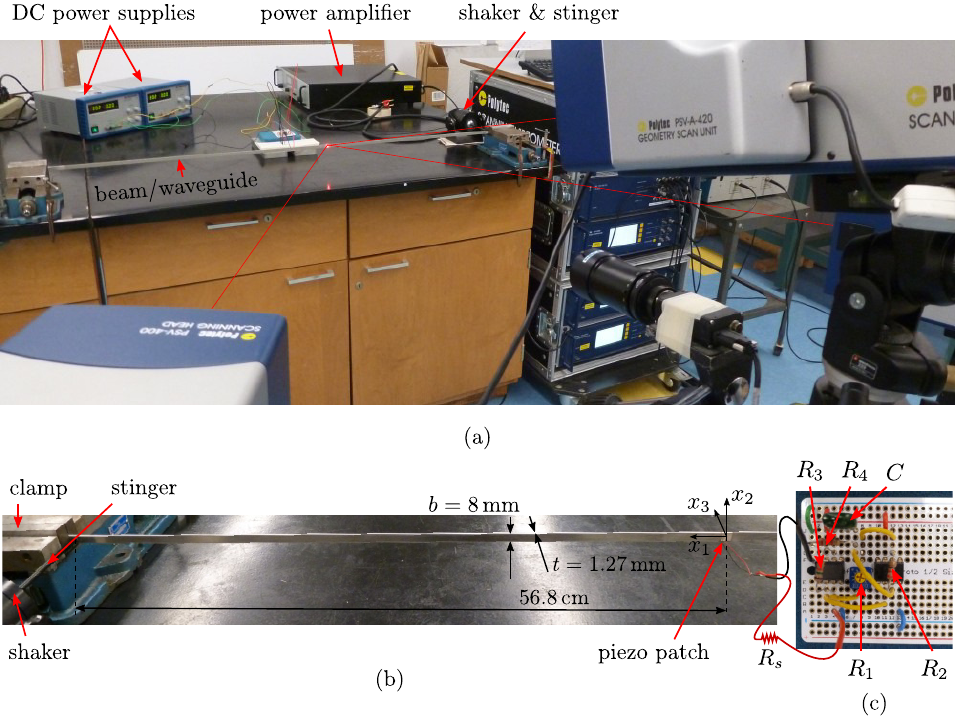}
\caption{(a) Experimental setup to investigate the interaction of a traveling wave with an electromechanical resonator (RL-shunted patch). (b) Detail of the beam/waveguide, with its characteristic dimensions; the beam features a single piezoelectric patch. (c) Synthetic inductor built on a solderable breadboard.}
\label{fig:exp1D}
\end{figure*}

The interaction between an elastic wave and an electromechanical resonator comprising a piezoelectric patch and a resistor-inductor shunt is investigated by means of a simple one-dimensional experiment. The setup is similar to the one used to study the behavior of the lattice, and is illustrated in Fig.~\ref{fig:exp1D}a.
The specimen is a $117.4\,\mathrm{cm}$ long and $8\,\mathrm{mm}$ wide beam carved out from the same 6061 Al sheet used to manufacture the lattice specimen (thickness $1.27\,\mathrm{mm}$). As illustrated in Fig.~\ref{fig:exp1D}b, a single piezoelectric patch (same as those used in the lattice experiment) is attached to the rear side of the beam, $56.8\,\mathrm{cm}$ away from the point where the stinger is attached. The circuit used in this experiment is slightly different from those used in the lattice case. In particular, all circuit parameters are identical except for $R_2$, which is $220\,\mathrm{\Omega}$ instead of $150\,\mathrm{\Omega}$. This circuit allows to tune the resonant frequency of the electromechanical resonator from $2.5\,\mathrm{kHz}$ to $16\,\mathrm{kHz}$.

\begin{figure*} [!htb]
\centering
\includegraphics[scale=1.45]{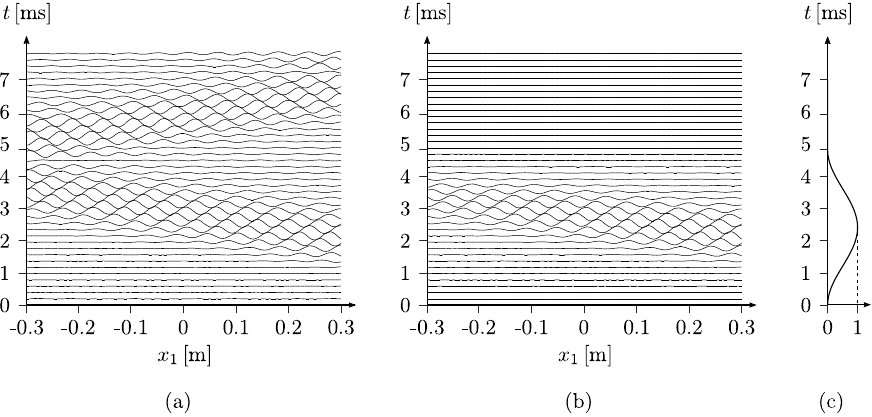}
\caption{Spatio-temporal evolution of a 9-cycle burst with carrier frequency $3.5\,\mathrm{kHz}$ within the measurement region of the beam. (a) Pristine signal, featuring a leftward-propagating packet and a rightward-propagating boundary reflection. (b) Filtered signal, without boundary reflection. (c) Time-domain filter used to obtain (b) from (a).}
\label{fig:filt}
\end{figure*}

The space-time evolution of a wave packet generated by a 9-cycle burst with carrier $3.5\mathrm{kHz}$, recorded along a portion of the beam spanning $30\,\mathrm{cm}$ to the left and to the right of the patch, is reported in Fig.~\ref{fig:filt}a.
The response presents two main features. For $t<4.5\,\mathrm{ms}$, we can see a leftward-propagating wave traveling from the source to the opposite clamp; for $t>4.5\,\mathrm{ms}$, we can see a packet propagating towards the source due to boundary reflections. In order to exclude the effects of the boundary from our data and to be able to treat this as a pure propagating wave problem (i.e. lifting the possibility for standing waves to be established in the beam due to interference of incident and reflected waves), we apply the time-domain filter shown in Fig.~\ref{fig:filt}c and obtain the filtered space-time data shown in Fig.~\ref{fig:filt}b. Since the separation between the two packets is not very pronounced, filtering will slightly affect the signal for values of $x_1$ towards the left of the domain.

\begin{figure} [!htb]
\centering
\includegraphics[scale=1.45]{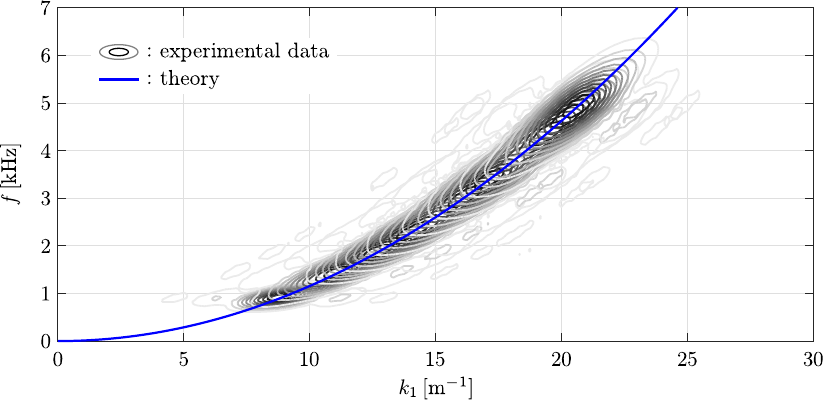}
\caption{Comparison between the experimentally reconstructed dispersion relation for the beam featuring a single piezoelectric patch (gray-to-black contours, where the dark color indicates larger spectral values) and the dispersion branch calculated from beam theory (superimposed red line).}
\label{fig:exp1Ddisp}
\end{figure}

To evaluate the influence of the piezoelectric patch in its open circuit state on the beam response, we probe the beam with multiple 5-cycle burst signals centered at 1, 1.5, 2, 2.5, 3, 3.5, 4, 4.5, 5 and $5.5\,\mathrm{kHz}$. In this case, we decided to use 5-cycle bursts to improve the separation between incident and reflected waves and to assure that the signals are not distorted by our filtering procedure. For each burst, we apply the 2D Discrete Fourier Transform to the recorded space-time data, and obtain a frequency versus wavenumber representation for the considered signal. By normalizing the response obtained for each burst and patching all the results together, we can reconstruct the dispersion relation as shown in Fig.~\ref{fig:exp1Ddisp}.
In this plot, the experimental data is reported as contours (there are 10 sets of contours, each one for one of the excitation signals mentioned above), where the darker lines correspond to larger values of the spectral amplitude $V_3$, which represents the Fourier transform of the out-of-plane velocity $v_3$. This reconstructed dispersion portrait is compared to the theoretical dispersion curve for an Euler-Bernoulli beam without piezoelectric patches, i.e.
\begin{equation}
\omega=k_1^2\sqrt{\frac{E I}{\rho A}}\,\,,
\label{eq:dispbeam}
\end{equation}
where the Young's modulus of 6061 aluminum is $E=68\,\mathrm{GPa}$, its density is $\rho=2700\,\mathrm{kg\,m^{-3}}$, the second moment of area is $I=(b\,t^3)/12$ and the cross-sectional area is $A=b\,t$ (the values given to $b$ and $t$ are those of the actual beam). The values of $k_1$ obtained from Eq.~\ref{eq:dispbeam} are then normalized by $2\pi$, to obtain units of $\mathrm{m^{-1}}$ consistent with those obtained from the experimental data. The reconstructed and calculated dispersion relations match well, and such agreement suggests that the effect of the physical presence of the piezoelectric patch on the wave response of the beam is negligible. This strengthens the claim made in the article that the electromechanical resonators obtained by shunting the piezo patch with RL circuits are minimally invasive when the shunting circuits are open, making the all-open configuration a fair representation of the default behavior of the pristine lattice.

\subsubsection*{A NUMERICAL DEMONSTRATION OF ANISOTROPY OVERRIDING}
In this Section, we provide a numerical demonstration of anisotropy overriding---the strategy for spatial wave manipulation in cellular periodic structures that we introduced in this work. The aim of this strategy is to leverage the interplay between the default anisotropically-propagating wave patterns of the cellular medium and the dynamics of resonators strategically placed at selected nodes of the lattice. Our starting point is the selection of a periodic cellular medium featuring anisotropic wave patterns at selected frequencies. We then consider a finite-size medium featuring the selected architecture and we introduce resonators in a subset of cells concentrated in certain subregion(s) of the lattice domain, in order to ``override'' only selected features of the anisotropic wave patterns. The main advantage of this approach is that a large number of resonators is not required to achieve significant wave focusing results.

\subsubsection*{A square lattice of springs and masses}
\begin{figure} [!htb]
\centering
\includegraphics[scale=1.45]{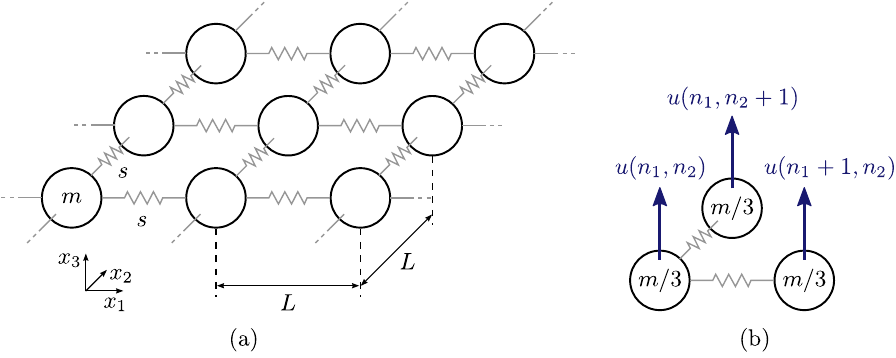}
\caption{Two-dimensional square lattice of springs and masses. (a) Lattice configuration. (b) Unit cell (the arrows indicate the degrees of freedom) used for the Bloch analysis.}
\label{fig:smuc}
\end{figure}

\begin{figure} [!htb]
\centering
\includegraphics[scale=1.45]{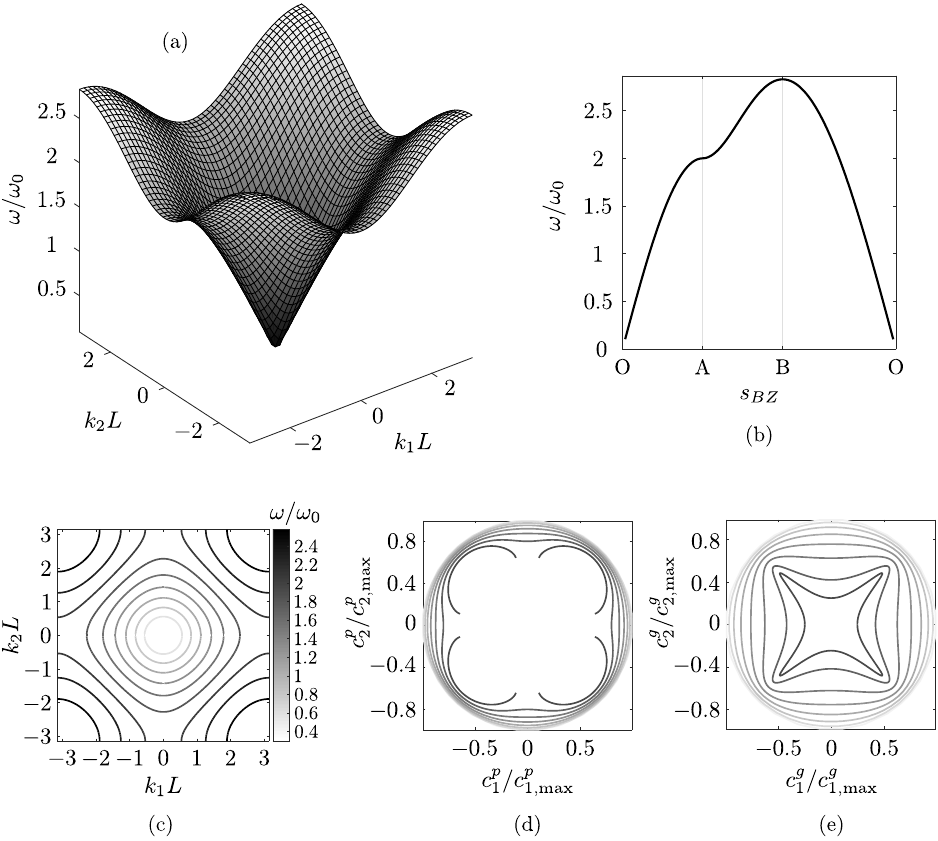}
\caption{Results of the unit cell analysis for a two-dimensional lattice of springs and masses. (a) Dispersion surface. (b) Band diagram. (c) Iso-frequency contours. (d) Phase velocity contours. (e) Group velocity contours.}
\label{fig:smucres}
\end{figure}

To illustrate the principles and potential of the \emph{anisotropy overriding} approach, we adopt the simplest two-dimensional periodic structure---a square lattice of springs and point masses. As shown in Fig.~\ref{fig:smuc}a, the lattice is uniform---all springs have stiffness $s$, all point masses have mass $m$ and are equally spaced (the inter-mass distance along both $x_1$ and $x_2$ is $L$).
Therefore, the position vector indicating a generic mass is $\mathbf{r}=n_1\,L\,\mathbf{i}_1+n_2\mathbf\,L\,\mathbf{i}_2$, where $n_1$ and $n_2$ are integer indices. As it is extensively done in the literature, we compute the wave response of this periodic structure through a Bloch-based unit cell analysis. The unit cell used for our calculations is illustrated in Fig.~\ref{fig:smuc}b. Considering values of mass of $m/3$ instead of $m$ is necessary to obtain the exact configuration shown in Fig.~\ref{fig:smuc}a, upon the application of Bloch's periodic boundary conditions. Each mass has a single out-of-plane degree of freedom, i.e. it can move vertically as illustrated by the arrows in Fig.~\ref{fig:smuc}b; the displacement of the $(n_1,n_2)$ mass is labeled $u(n_1,n_2)$. Hence, the springs have to be intended as constraining the vertical motion of the masses. As a result, upon application of Bloch's conditions, the characteristic equation leads to a mono-branch dispersion relation which corresponds to a single mode of wave propagation:
\begin{equation}
\omega=\sqrt{\frac{2s}{m}\left( 2-\cos{k_1L}-\cos{k_2L} \right)}\,\,,
\label{eq:smuc}
\end{equation}
where $\omega$ is the angular frequency, while $k_1$ and $k_2$ are the Cartesian components of the wave vector. The dispersion surface for parameters $s=m=L=1$ is shown in Fig.~\ref{fig:smucres}a, where we used a normalization factor $\omega_0=\sqrt{s/m}$ for $\omega$, and $L$ for $k_1$ and $k_2$. 
The response is more compactly represented by the band diagram in Fig.~\ref{fig:smucres}b, where $s_{\mathrm{BZ}}$ is a coordinate spanning the contour of the IBZ of the lattice. In the reciprocal lattice coordinate system, the coordinates of the corners of the IBZ are O $(0,0)$, A $(0,\pi)$, B $(\pi,\pi)$. As the frequency increases, we record a transition from isotropic to anisotropic wave behavior. This aspect is clearly indicated by the iso-frequency contours of the dispersion surface, shown in Fig.~\ref{fig:smucres}c, and by the phase and group velocity contours in Fig.~\ref{fig:smucres}d and Fig.~\ref{fig:smucres}e, respectively. This anisotropy is correlated to the arising of partial bandgaps along the $x_1$ and $x_2$ directions (denoted by the absence of points along those directions for some of the iso-frequency contours), which cause the waves to predominantly propagate along $\pm 45^{\mathrm{o}}$-oriented directions for frequencies greater than $\omega\approx 1.8\,\omega_0$.

\subsubsection*{A square lattice of springs, masses and resonators}
\begin{figure} [!htb]
\centering
\includegraphics[scale=1.45]{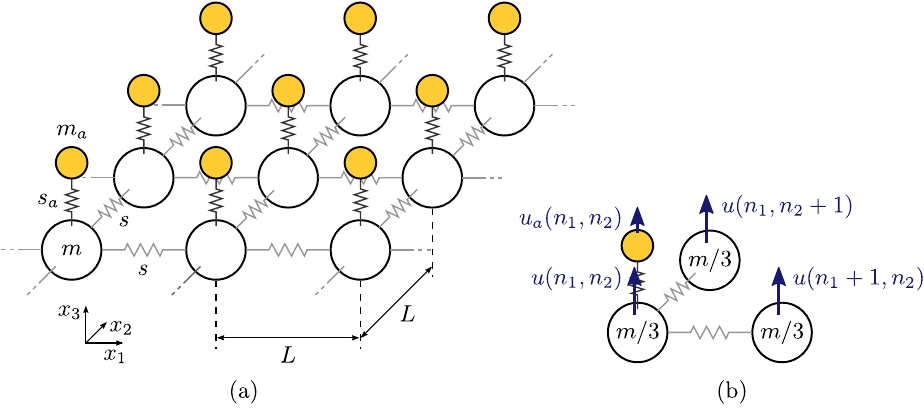}
\caption{Two-dimensional square lattice of springs and masses, with auxiliary resonators (yellow masses). (a) Lattice configuration. (b) Unit cell (the arrows indicate the degrees of freedom) used for the Bloch analysis.}
\label{fig:smauc}
\end{figure}

\begin{figure} [!htb]
\centering
\includegraphics[scale=1.45]{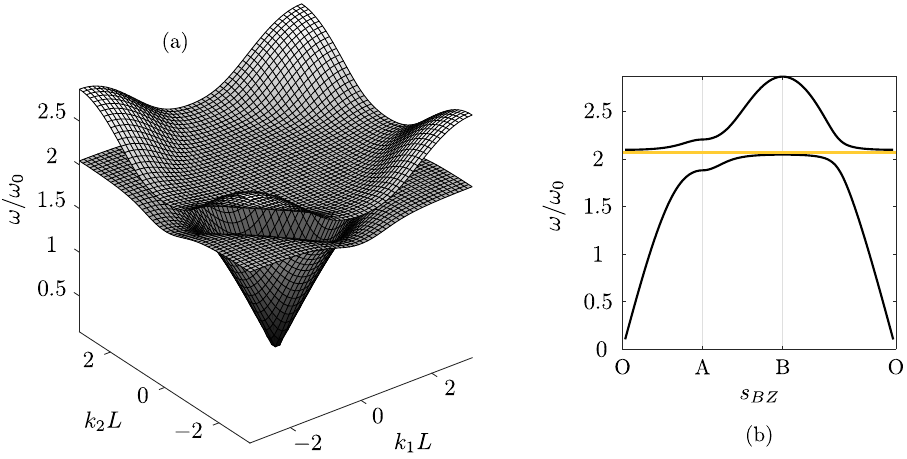}
\caption{Results of the unit cell analysis for a two-dimensional lattice of springs and masses, featuring out-of-plane spring-mass resonators. (a) Dispersion surfaces. (b) Band diagram (the yellow region indicates the resonator-induced bandgap).}
\label{fig:smaucres}
\end{figure}

To understand the influence of localized resonators on the wave response of the spring-mass lattice introduced in the previous Section, we consider a variation of the same lattice comprising an out-of-plane spring-mass resonator for each mass. This simple model is similar in nature to the one-dimensional ``mass-in-mass'' one introduced by Huang et al. [H. H. Huang et al, \emph{Int. J. Eng. Sci.} {\bf 47}(4), 610--617 (2009)]. All resonators in the periodic lattice are characterized by stiffness $s_a$ and mass $m_a$. The lattice is sketched in Fig.~\ref{fig:smauc}a, while the unit cell with all its degrees of freedom is shown in Fig.~\ref{fig:smauc}b. 
To preserve the characteristics of the original structure, the unit cell comprises a single resonator located at the $(n_1,n_2)$ mass. Upon the application of Bloch's conditions, the response of the infinite lattice reduces to a two degree of freedom problem. The two resulting dispersion surfaces can be computed by solving the following equation for $\omega$:
\begin{equation}
\mathrm{det}\left[\!\!\!\begin{array}{c c}
s_a+2s\left( 2-\cos{k_1L}-\cos{k_2L} \right)-\omega^2m & -s_a\\
-s_a & s_a-\omega^2m_a \end{array}\,\,\,\right]=0
\label{eq:smauc}
\end{equation}
and by selecting only the positive roots. The dispersion surface obtained with parameters $s=m=1$, $s_a/s=0.1$ and $m_a/m=0.0233$ is shown in Fig.~\ref{fig:smaucres}a. 
The band diagram in Fig.~\ref{fig:smaucres}b features a thin locally-resonant bandgap, highlighted in yellow, whose onset matches the resonant frequency of the resonator $\sqrt{s_a/m_a}=2.0717\,\omega_0$. The parameters of the resonator are chosen as to induce a locally resonant bandgap in a frequency region where the response of the lattice without resonators is strongly anisotropic (recall that, from Fig.~\ref{fig:smucres}, anisotropy is observed above $1.8\,\omega_0$). This choice is motivated by the fact that we want our resonators to interact with the anisotropic wave patterns inherent to the baseline architecture without resonators. 

\subsubsection*{Anisotropy overriding: partial bandgaps through strategically-located resonators}
\begin{figure} [!htb]
\centering
\includegraphics[scale=1.45]{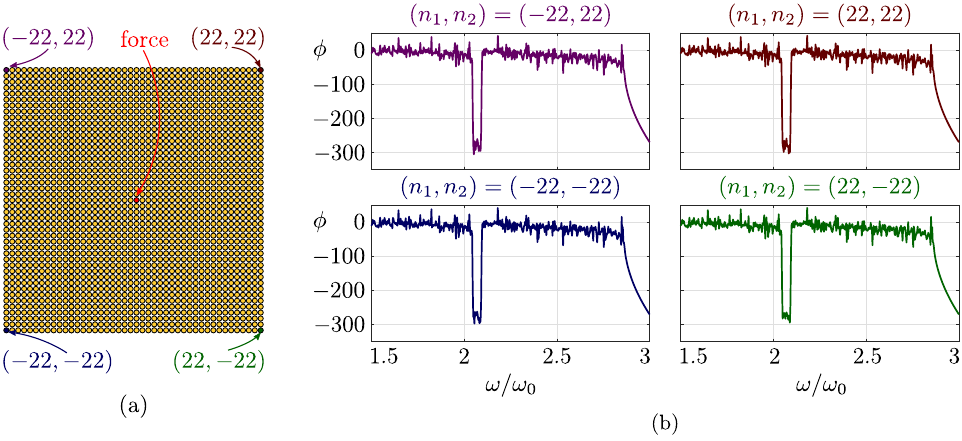}
\caption{Steady state response of a lattice in which all masses are connected to resonators. (a) Illustration of the square lattice structure. In this representation, yellow masses feature resonators, while white ones do not. The red point indicates the node where the force is applied. The corner nodes are color coded. (b) Transmissibilities recorded at the four corner nodes of the lattice.}
\label{fig:smafssh1}
\end{figure}

\begin{figure} [!htb]
\centering
\includegraphics[scale=1.45]{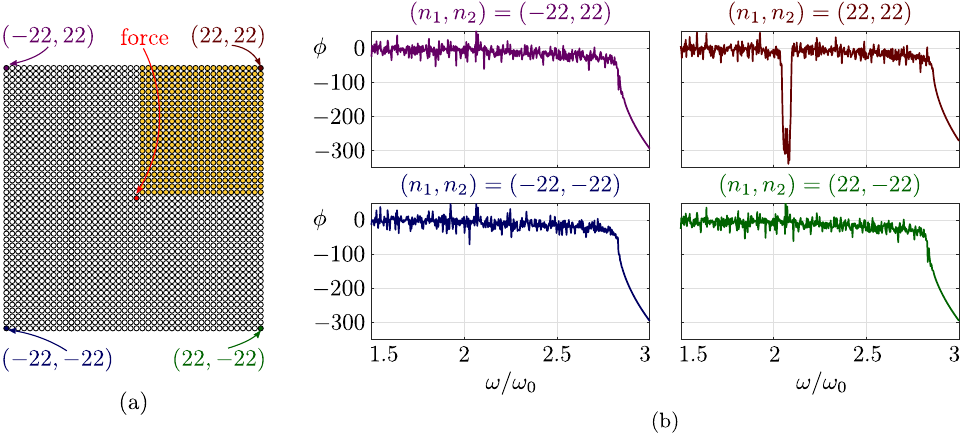}
\caption{Steady state response of a lattice with resonators located in the top-right quadrant only. (a) Illustration of the square lattice structure. In this representation, yellow masses feature resonators, while white ones do not. The red point indicates the node where the force is applied. The corner nodes are color coded. (b) Transmissibilities recorded at the four corner nodes of the lattice.}
\label{fig:smafssh2}
\end{figure}

We now have all the ingredients necessary to introduce our spatial wave manipulation strategy, which is inherently applicable only to finite-sized structures. To do so, we consider an array of 45$\times$45 masses and we place resonators on selected masses. Throughout this analysis, the excitation is always applied at the center of the lattice, i.e., at the point corresponding to $(n_1,n_2)=(0,0)$, while the response is sensed at the four corners of the lattice, i.e. at $(n_1,n_2)=(-22,22)$, $(n_1,n_2)=(22,-22)$, $(n_1,n_2)=(22,22)$ and $(n_1,n_2)=(-22,22)$. These characteristic points are color coded in the sketch of the lattice in Fig.~\ref{fig:smafssh1}a. In this sketch, and in the following ones in this Section, the yellow masses feature resonators while the white ones do not.
Fig.~\ref{fig:smafssh1}b represents the steady state response of the lattice with omnipresent resonators shown in Fig.~\ref{fig:smafssh1}a. The response is given in terms of transmissibilities obtained by normalizing the displacement recorded at the four corners of the lattice by the displacement of the excitation point. As expected, all four corners display a bandgap in their transmissibility; this attenuation region matches the one observed in the unit cell analysis of the same configuration (Fig.~\ref{fig:smaucres}b). Note that the cut-off observed at $\omega\approx 2.8\,\omega_0$, which is consistent with the absence of modes above that frequency in the dispersion relation, is due to the discrete nature of our model (which has a finite number of degrees of freedom). In Fig.~\ref{fig:smafssh2}, on the other hand, we report the response of a lattice where the only masses featuring resonators (yellow masses in Fig.~\ref{fig:smafssh2}a) are located in the top-right quadrant. 
As shown in Fig.~\ref{fig:smafssh2}b, this modification of the lattice causes the appearance of a partial bandgap along the direction connecting the excitation point to the $(22,22)$ mass, while the responses along the other directions are not significantly affected by the presence of the resonators. Clearly, placing resonators in other sub-regions would produce partial bandgaps along other directions.

\begin{figure} [!htb]
\centering
\includegraphics[scale=1.45]{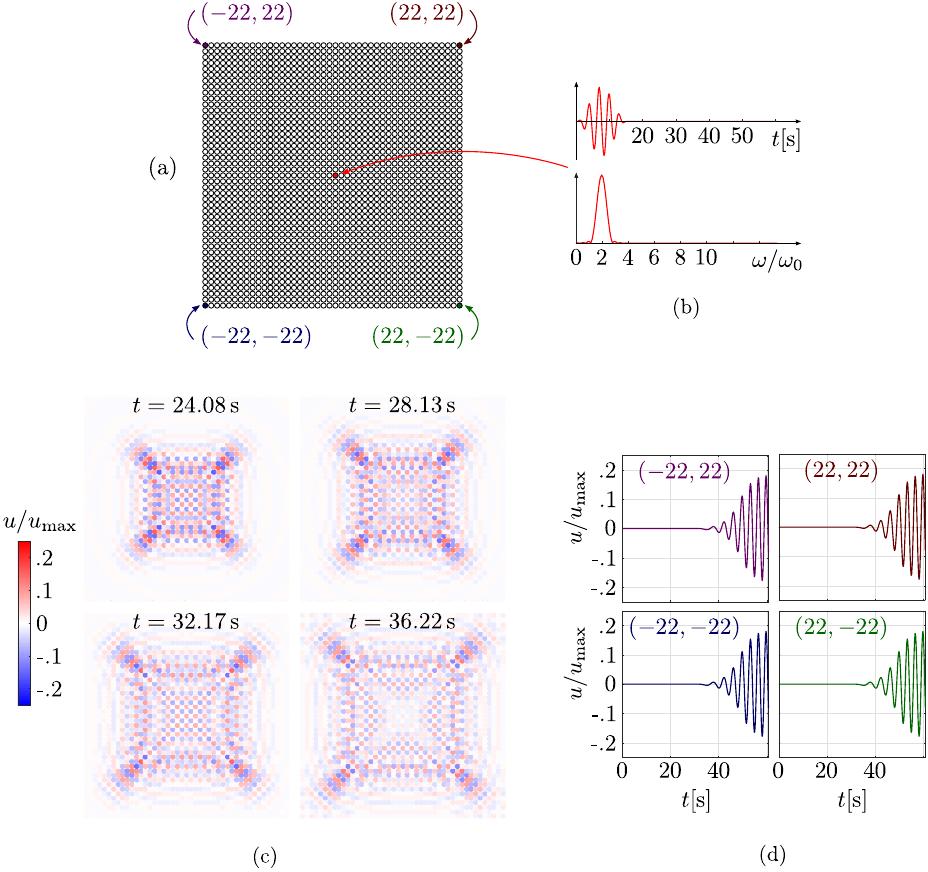}
\caption{Transient response of a square lattice of springs and masses, without resonators. (a) Lattice configuration with color-coded key points. (b) Input signal---5 cycle burst with carrier frequency $\Omega=2.07\,\omega_0$. (c) Displacement wavefields at four time instants. (d) Time histories at the four corners of the lattice.}
\label{fig:smafsst1}
\end{figure}

\begin{figure} [!htb]
\centering
\includegraphics[scale=1.45]{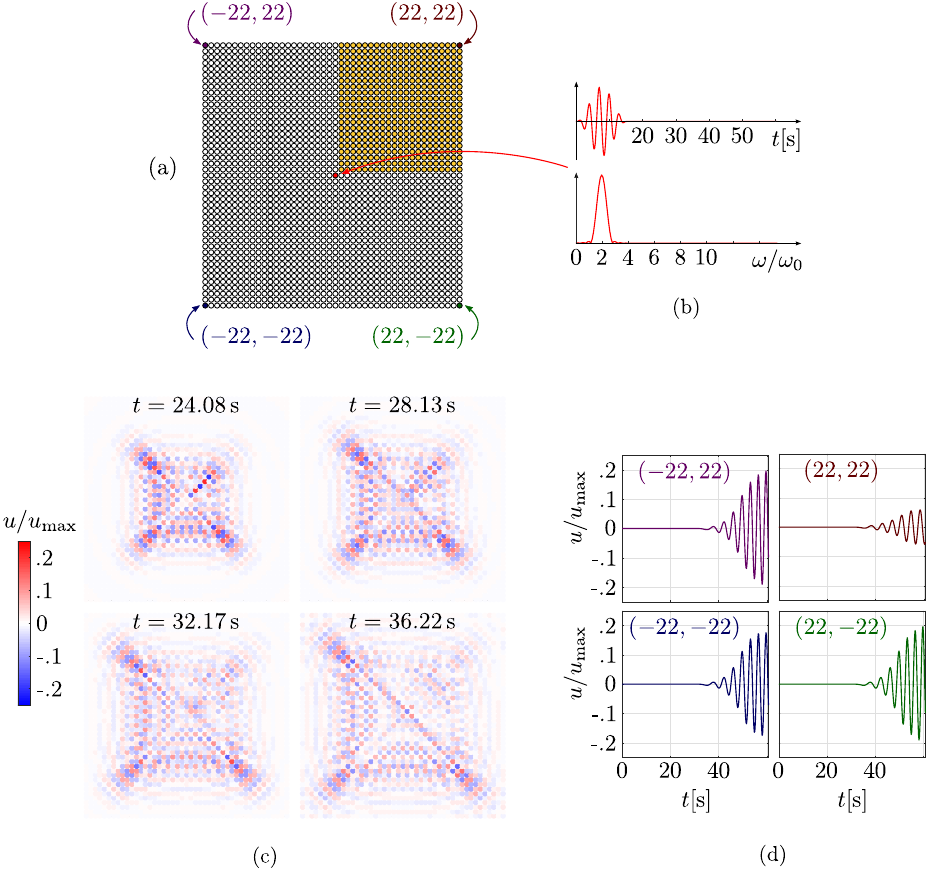}
\caption{Transient response of a square lattice of springs and masses, with resonators connected to the masses belonging to the top-right quadrant of the domain. (a) Lattice configuration with color-coded key points. (b) Input signal---5 cycle burst with carrier frequency $\Omega=2.07\,\omega_0$. (c) Displacement wavefields at four time instants. (d) Time histories at the four corners of the lattice.}
\label{fig:smafsst2}
\end{figure}

Our strategy relies on the mutual interaction between the inherent anisotropy of the response of a periodic medium and the partial bandgaps generated by strategically placing resonators as in Fig.~\ref{fig:smafssh2}. To harness the strongest anisotropy, we excite the lattice with a burst signal with carrier frequency $2.07\,\omega_0$. To maximize the influence of the resonators on the transient wave, and recalling that the onset of a locally-resonant bandgap coincides with the resonators' natural frequency, we tune them at the carrier frequency of the burst. Recall, from Fig.~\ref{fig:smucres}, that this frequency corresponds to a propagation pattern with beaming along $\pm 45^{\mathrm{o}}$-oriented directions for a lattice without resonators. The response of a 45$\times$45 pristine lattice, shown in Fig.~\ref{fig:smafsst1}a, to a signal whose time evolution and frequency spectrum are shown in Fig.~\ref{fig:smafsst1}b, is reported in Figs.~\ref{fig:smafsst1}c-d.
The wavefields show an ``x-shaped'' response that is consistent with the unit cell analysis prediction, and their main feature is represented by four spatially confined packets propagating along the $\pm 45^{\mathrm{o}}$-oriented directions. Our goal is to remove one (or multiple) of these packets through the introduction of resonators and the activation of partial bandgaps. An example of the application of this strategy is shown in Fig.~\ref{fig:smafsst2}, where all the masses in the top-right quadrant of the domain feature resonators. 
The wavefields in Fig.~\ref{fig:smafsst2}c and the time histories at the corner nodes in Fig.~\ref{fig:smafsst2}d show that, as expected, the top-right portion of the domain is significantly de-energized by the presence of the resonators. As a result, one of the four original packets is attenuated and the wave response loses its two-fold symmetry. Note that other wave focusing scenarios can be obtained by introducing resonators in other portions of the domain. It is worth pointing out that the amount of distinct wave focusing effects achievable with this strategy is symbiotic to the richness of anisotropic characteristics of the considered cellular medium. In the case of the mono-modal spring mass lattice discussed in this section, the wave focusing scenarios that can be produced are inevitably limited.

\subsubsection*{Minimizing the number of resonators}
\begin{figure} [!htb]
\centering
\includegraphics[scale=1.45]{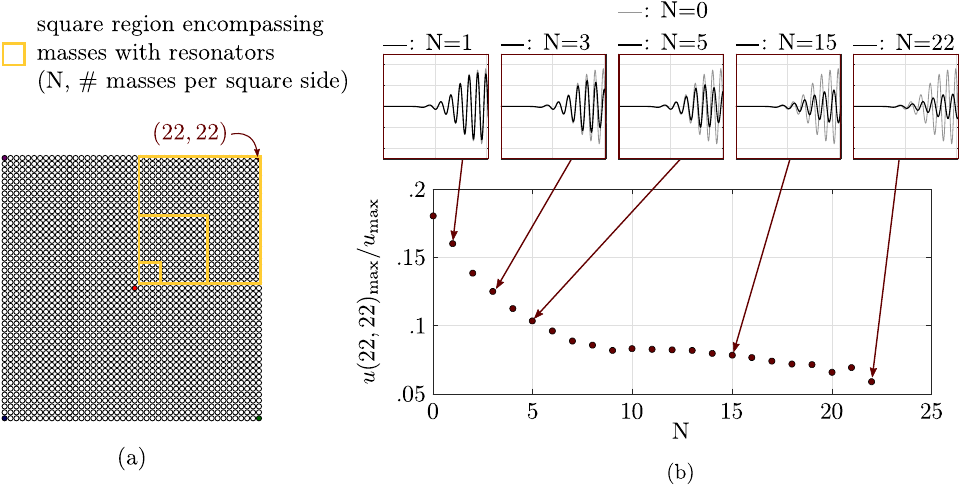}
\caption{Dependence of anisotropy overriding on the number of resonators. (a) Lattice featuring a varying number of resonators arranged within a square portion of the domain comprising N masses per side and having bottom-left corner at the $(1,1)$ mass. (b) Maximum displacement recorded at the $(22,22)$ corner, versus N. The black lines in the inserts show the wave responses for selected numbers of masses, while the baseline gray lines represent the case without resonators.}
\label{fig:smafsst4}
\end{figure}

So far, we have shown the effectiveness of the anisotropy overriding strategy in the context of lattice structures where entire quadrants of the domain feature resonators. Deploying large numbers of resonators may, however, be impractical in experimental applications. With these constraints in mind, we are now interested in understanding how the number of resonators impacts the achievable symmetry modification of the anisotropic patterns, thus determining some lower bounds that can be valuable in designing effective---yet parsimonious---wave control strategies. To carry out this parametric analysis, we only introduce resonators in a square portion of the domain (with varying number of masses per side labeled by N) whose bottom left corner is fixed to the $(1,1)$ mass. Some of the considered square sub-domains are highlighted by yellow boxes in Fig.~\ref{fig:smafsst4}.
The comparison between different configurations is carried out by plotting the maximum displacement recorded at the top-right corner, $u_{\mathrm{max}}(22,22)$ (normalized by the maximum displacement recorded in the whole domain, $u_{\mathrm{max}}$) against the number of masses per each side of the square subdomain featuring resonators. This trend is illustrated in Fig.~\ref{fig:smafsst4}b. For N smaller than 10, the amplitude of the displacement decreases exponentially with N, while for N greater than 10, the decrease is much less pronounced. This suggests that introducing more than 10$\times$10 resonators does not produce any significant advantage in terms of attenuation. Most importantly, the exponential nature of the first part of the curve highlights how we can obtain significant attenuation effects and therefore obtain the desired wave patterns even with a limited number of resonators.

\subsubsection*{ADDITIONAL RESULTS ON ANISOTROPY OVERRIDING}
\begin{figure*} [!t]
\centering
\includegraphics[scale=1.45]{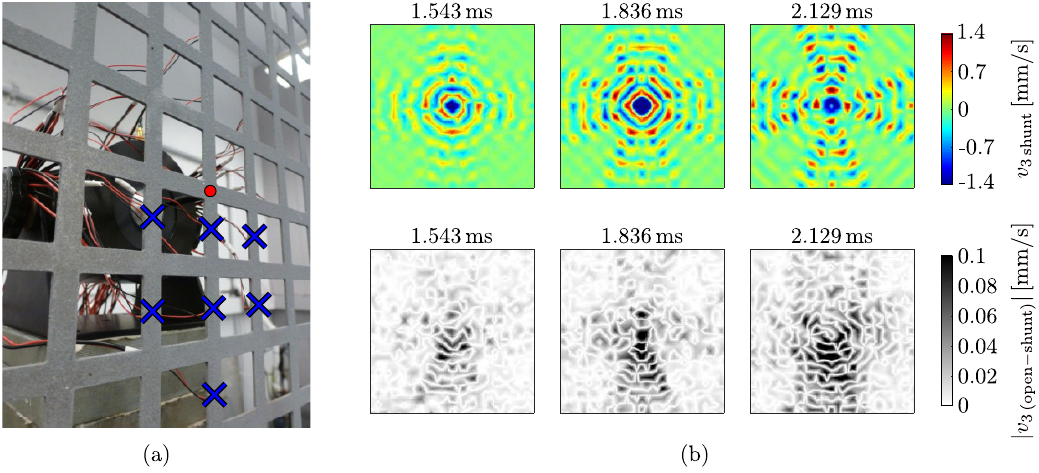}
\caption{Anisotropy overriding at $4.1\,\mathrm{kHz}$; the targeted wave feature is the downward-propagating packet. (a) Detail of the front (scanned) face of the specimen, indicating where the shunted patches are located. (b) Top row: wavefields acquired at three time instants. Bottom row: spatial patterns displayed by the difference between the open circuit and the shunted response.}
\label{fig:resB}
\end{figure*}

In this Section, we report one last anisotropy overriding scenario. As in Fig.~8, we consider the response to a 13-cycle burst with carrier frequency $4.1\,\mathrm{kHz}$. We shunt the patches belonging to the marked links in Fig.~\ref{fig:resB}a and, as shown in Fig.~\ref{fig:resB}b, we alter the packet propagating downwards from the excitation location.


{\small
\begin{thebibliography}{10}

\bibitem{Fleck_PRSA_2010}
N.~A. Fleck, V.~S. Deshpande, and M.~F. Ashby.
\newblock Micro-architectured materials: past, present and future.
\newblock {\em Proc. R. Soc. A}, 466(2121):2495--2516, 2010.

\bibitem{Schaedler_SCIENCE_2011}
T.~A. Schaedler, A.~J. Jacobsen, A.~Torrents, A.~E. Sorensen, J.~Lian, J.~R.
  Greer, L.~Valdevit, and W.~B. Carter.
\newblock Ultralight metallic microlattices.
\newblock {\em Science}, 334(6058):962--965, 2011.

\bibitem{Zheng_SCIENCE_2014}
X.~Zheng, H.~Lee, T.~H. Weisgraber, M.~Shusteff, J.~DeOtte, E.~B. Duoss, J.~D.
  Kuntz, M.~M. Biener, Q.~Ge, J.~A. Jackson, S.~O. Kucheyev, N.~X. Fang, and
  C.~M. Spadaccini.
\newblock Ultralight, ultrastiff mechanical metamaterials.
\newblock {\em Science}, 344(6190):1373--1377, 2014.

\bibitem{Meza_PNAS_2015}
L.~R. Meza, A.~J. Zelhofer, N.~Clarke, A.~J. Mateos, D.~M. Kochmann, and J.~R.
  Greer.
\newblock Resilient 3d hierarchical architected metamaterials.
\newblock {\em Proc. Natl. Acad. Sci. U.S.A.}, 112(37):11502--11507, 2015.

\bibitem{Coulais_NATURE_2016}
C.~Coulais, E.~Teomy, K.~de~Reus, Y.~Shokef, and M.~van Hecke.
\newblock Combinatorial design of textured mechanical metamaterials.
\newblock {\em Nature}, 535:529--532, 2016.

\bibitem{Shan_ADMA_2015}
S.~Shan, S.~H. Kang, J.~R. Raney, P.~Wang, L.~Fang, F.~Candido, J.~A. Lewis,
  and K.~Bertoldi.
\newblock Multistable architected materials for trapping elastic strain energy.
\newblock {\em Adv. Mater.}, 27(29):4296--4301, 2015.

\bibitem{Restrepo_EML_2015}
D.~Restrepo, N.~D. Mankame, and P.~D. Zavattieri.
\newblock Phase transforming cellular materials.
\newblock {\em Extreme Mech. Lett.}, 4:52--60, 2015.

\bibitem{Buckmann_NATCOMM_2014}
T.~B\"{u}ckmann, M.~Thiel, M.~Kadic, R.~Schittny, and M.~Wegener.
\newblock An elasto-mechanical unfeelability cloak made of pentamode
  metamaterials.
\newblock {\em Nat. Commun.}, 5:4130, 2014.

\bibitem{Lakes_ADMA_1993}
R.~Lakes.
\newblock Advances in negative poisson's ratio materials.
\newblock {\em Adv. Mater.}, 5(4):293--296, 1993.

\bibitem{Wang_PRL_2016}
Q.~Wang, J.~A. Jackson, Q.~Ge, J.~B. Hopkins, C.~M. Spadaccini, and N.~X. Fang.
\newblock Lightweight mechanical metamaterials with tunable negative thermal
  expansion.
\newblock {\em Phys. Rev. Lett.}, 117(17):175901, 2016.

\bibitem{Song_ADEM_2016}
Y.~Song, P.~C. Dohm, B.~Haghpanah, A.~Vaziri, and J.~B. Hopkins.
\newblock An active microarchitectured material that utilizes piezo actuators
  to achieve programmable properties.
\newblock {\em Adv. Eng. Mater.}, 18(7):1113--1117, 2016.

\bibitem{Kroedel_PRAPP_2016}
S.~Kr\"odel and C.~Daraio.
\newblock Microlattice metamaterials for tailoring ultrasonic transmission with
  elastoacoustic hybridization.
\newblock {\em Phys. Rev. Applied}, 6(6):064005, 2016.

\bibitem{Phani_JASA_2006}
A.~S. Phani, J.~Woodhouse, and N.~A. Fleck.
\newblock Wave propagation in two-dimensional periodic lattices.
\newblock {\em J. Acoust. Soc. Am.}, 119(4):1995--2005, 2006.

\bibitem{Baravelli_JSV_2013}
E.~Baravelli and M.~Ruzzene.
\newblock Internally resonating lattices for bandgap generation and
  low-frequency vibration control.
\newblock {\em J. Sound Vib.}, 332(25):6562--6579, 2013.

\bibitem{Kroedel_AEM_2014}
S.~Kr\"odel, T.~Delpero, A.~Bergamini, P.~Ermanni, and D.~M. Kochmann.
\newblock 3d auxetic microlattices with independently controllable acoustic
  band gaps and quasi-static elastic moduli.
\newblock {\em Adv. Eng. Mater.}, 16(4):357--363, 2014.

\bibitem{Junyi_IJSS_2016}
L.~Junyi and D.S. Balint.
\newblock A parametric study of the mechanical and dispersion properties of
  cubic lattice structures.
\newblock {\em Int. J. Solids Struct.}, 91:55--71, 2016.

\bibitem{Miniaci_APL_2016}
M.~Miniaci, A.~Krushynska, A.~B. Movchan, F.~Bosia, and N.~M. Pugno.
\newblock Spider web-inspired acoustic metamaterials.
\newblock {\em Appl. Phys. Lett.}, 109(7):071905, 2016.

\bibitem{Matlack_PNAS_2016}
K.~H. Matlack, A.~Bauhofer, S.~Kr\"{o}del, A.~Palermo, and C.~Daraio.
\newblock Composite 3d-printed metastructures for low-frequency and broadband
  vibration absorption.
\newblock {\em Proc. Natl. Acad. Sci. U.S.A.}, 113(30):8386--8390, 2016.

\bibitem{Warmuth_SCIREP_2017}
F.~Warmuth, M.~Wormser, and C.~K\"{o}rner.
\newblock Single phase 3d phononic band gap material.
\newblock {\em Sci. Rep.}, 7:3843, 2017.

\bibitem{Chernow_APL_2015}
V.~F. Chernow, H.~Alaeian, J.~A. Dionne, and J.~R. Greer.
\newblock Polymer lattices as mechanically tunable 3-dimensional photonic
  crystals operating in the infrared.
\newblock {\em Appl. Phys. Lett.}, 107(10):101905, 2015.

\bibitem{Langley_JSV_1996}
R.~S. Langley.
\newblock The response of two-dimensional periodic structures to point harmonic
  forcing.
\newblock {\em J. Sound. Vib.}, 197(4):447--469, 1996.

\bibitem{Ruzzene_SMS_2003}
M.~Ruzzene, F.~Scarpa, and F.~Soranna.
\newblock Wave beaming effects in two-dimensional cellular structures.
\newblock {\em Smart Mater. Struct.}, 12(3):363, 2003.

\bibitem{Wen_JPD_2008}
J.~Wen, D.~Yu, G.~Wang, and X.~Wen.
\newblock Directional propagation characteristics of flexural wave in
  two-dimensional periodic grid-like structures.
\newblock {\em J. Phys. D: Appl. Phys.}, 41(13):135505, 2008.

\bibitem{Carta_IJSS_2014}
G.~Carta, M.~Brun, A.B. Movchan, N.V. Movchan, and I.S. Jones.
\newblock Dispersion properties of vortex-type monatomic lattices.
\newblock {\em Int. J. Solids Struct.}, 51(11):2213--2225, 2014.

\bibitem{Celli_JSV_2014}
P.~Celli and S.~Gonella.
\newblock Laser-enabled experimental wavefield reconstruction in
  two-dimensional phononic crystals.
\newblock {\em J. Sound. Vib.}, 333(1):114--123, 2014.

\bibitem{Wang_JPD_2014}
Y-F. Wang, Y-S. Wang, and C.~Zhang.
\newblock Bandgaps and directional propagation of elastic waves in 2d square
  zigzag lattice structures.
\newblock {\em J. Phys. D: Appl. Phys.}, 47(48):485102, 2014.

\bibitem{Trainiti_IJSS_2016}
G.~Trainiti, J.J. Rimoli, and M.~Ruzzene.
\newblock Wave propagation in undulated structural lattices.
\newblock {\em Int. J. Solids Struct.}, 97:431--444, 2016.

\bibitem{Zelhofer_IJSS_2017}
A.~J. Zelhofer and D.~M. Kochmann.
\newblock On acoustic wave beaming in two-dimensional structural lattices.
\newblock {\em Int. J. Solids Struct.}, 115-116:248--269, 2017.

\bibitem{Ganesh_APL_2017}
R.~Ganesh and S.~Gonella.
\newblock Experimental evidence of directivity-enhancing mechanisms in
  nonlinear lattices.
\newblock {\em Appl. Phys. Lett.}, 110(8):084101, 2017.

\bibitem{Lefebvre_PRL_2017}
G.~Lefebvre, T.~Antonakakis, Y.~Achaoui, R.~V. Craster, S.~Guenneau, and
  P.~Sebbah.
\newblock Unveiling extreme anisotropy in elastic structured media.
\newblock {\em Phys. Rev. Lett.}, 118(25):254302, 2017.

\bibitem{Celli_JAP_2014}
P.~Celli and S.~Gonella.
\newblock Low-frequency spatial wave manipulation via phononic crystals with
  relaxed cell symmetry.
\newblock {\em J. Appl. Phys.}, 115(10):103502, 2014.

\bibitem{Krattiger_AIP_2016}
D.~Krattiger, R.~Khajehtourian, C.~L. Bacquet, and M.~I. Hussein.
\newblock Anisotropic dissipation in lattice metamaterials.
\newblock {\em AIP Adv.}, 6(12):121802, 2016.

\bibitem{Ruzzene_JVA_2000}
M.~Ruzzene and A.~Baz.
\newblock Control of wave propagation in periodic composite rods using shape
  memory inserts.
\newblock {\em J. Vib. Acoust.}, 122(2):151--159, 2000.

\bibitem{Shan_ADFMA_2014}
S.~Shan, S.~H. Kang, P.~Wang, C.~Qu, S.~Shian, E.~R. Chen, and K.~Bertoldi.
\newblock Harnessing multiple folding mechanisms in soft periodic structures
  for tunable control of elastic waves.
\newblock {\em Adv. Funct. Mater.}, 24(31):4935--4942, 2014.

\bibitem{Nouh_JIMSS_2016}
M.~A. Nouh, O.~J. Aldraihem, and A.~Baz.
\newblock Periodic metamaterial plates with smart tunable local resonators.
\newblock {\em J. Intell. Mater. Syst. Struct.}, 27(13):1829--1845, 2016.

\bibitem{Zhu_APL_2016}
R.~Zhu, Y.~Y. Chen, M.~V. Barnhart, G.~K. Hu, C.~T. Sun, and G.~L. Huang.
\newblock Experimental study of an adaptive elastic metamaterial controlled by
  electric circuits.
\newblock {\em Appl. Phys. Lett.}, 108(1):011905, 2016.

\bibitem{Zhu_JASA_2016}
R.~Zhu, Y.~Y. Chen, Y.~S. Wang, G.~K. Hu, and G.~L. Huang.
\newblock A single-phase elastic hyperbolic metamaterial with anisotropic mass
  density.
\newblock {\em J. Acoust. Soc. Am.}, 139(6):3303--3310, 2016.

\bibitem{Croenne_JASA_2016}
C.~Cro\"{e}nne, M.~Ponge, B.~Dubus, C.~Granger, L.~Haumesser, F.~Levassort,
  J.~O. Vasseur, A.~Lordereau, M.~P. Thi, and A.~C. Hladky-Hennion.
\newblock Tunable phononic crystals based on piezoelectric composites with 1-3
  connectivity.
\newblock {\em J. Acoust. Soc. Am.}, 139(6):3296--3302, 2016.

\bibitem{Wang_ADMA_2016}
Z.~Wang, Q.~Zhang, K.~Zhang, and G.~K. Hu.
\newblock Tunable digital metamaterial for broadband vibration isolation at low
  frequency.
\newblock {\em Adv. Mater.}, 28(44):9857--9861, 2016.

\bibitem{Ouisse_SMS_2016}
M.~Ouisse, M.~Collet, and F.~Scarpa.
\newblock A piezo-shunted kirigami auxetic lattice for adaptive elastic wave
  filtering.
\newblock {\em Smart Mater. Struct.}, 25(11):115016, 2016.

\bibitem{Chen_PRAPP_2017}
Y.~Chen, T.~Li, F.~Scarpa, and L.~Wang.
\newblock Lattice metamaterials with mechanically tunable poisson's ratio for
  vibration control.
\newblock {\em Phys. Rev. Applied}, 7(2):024012, 2017.

\bibitem{Bilal_ADVMAT_2017}
O.~R. Bilal, A.~Foehr, and C.~Daraio.
\newblock Reprogrammable phononic metasurfaces.
\newblock {\em Adv. Mater.}, 29(39):1700628, 2017.

\bibitem{Celli_APL_2015}
P.~Celli and S.~Gonella.
\newblock Tunable directivity in metamaterials with reconfigurable cell
  symmetry.
\newblock {\em Appl. Phys. Lett.}, 106(9):091905, 2015.

\bibitem{Liu_SCIENCE_2000}
Z.~Liu, X.~Zhang, Y.~Mao, Y.~Y. Zhu, Z.~Yang, C.~T. Chan, and P.~Sheng.
\newblock Locally resonant sonic materials.
\newblock {\em Science}, 289(5485):1734--1736, 2000.

\bibitem{Cardella_SMS_2016}
D.~Cardella, P.~Celli, and S.~Gonella.
\newblock Manipulating waves by distilling frequencies: a tunable shunt-enabled
  rainbow trap.
\newblock {\em Smart Mater. Struct.}, 25(8):085017, 2016.

\bibitem{Lemoult_NATPHYS_2013}
F.~Lemoult, N.~Kaina, M.~Fink, and G.~Lerosey.
\newblock Wave propagation control at the deep subwavelength scale in
  metamaterials.
\newblock {\em Nat. Phys.}, 9(1):55--60, 2013.

\bibitem{Hagood_JSV_1991}
N.~W. Hagood and A.~von Flotow.
\newblock Damping of structural vibrations with piezoelectric materials and
  passive electrical networks.
\newblock {\em J. Sound Vib.}, 146(2):243--268, 1991.

\bibitem{Casadei_SMS_2010}
F.~Casadei, M.~Ruzzene, L.~Dozio, and K.~A. Cunefare.
\newblock Broadband vibration control through periodic arrays of resonant
  shunts: experimental investigation on plates.
\newblock {\em Smart Mater. Struct.}, 19(1):015002, 2010.

\bibitem{Airoldi_NJoP_2011}
L.~Airoldi and M.~Ruzzene.
\newblock Design of tunable acoustic metamaterials through periodic arrays of
  resonant shunted piezos.
\newblock {\em New J. Phys.}, 13(11):113010, 2011.

\bibitem{Casadei_JAP_2012}
F.~Casadei, T.~Delpero, A.~Bergamini, P.~Ermanni, and M.~Ruzzene.
\newblock Piezoelectric resonator arrays for tunable acoustic waveguides and
  metamaterials.
\newblock {\em J. Appl. Phys.}, 112(6):064902, 2012.

\bibitem{Bergamini_ADMA_2014}
A.~Bergamini, T.~Delpero, L.~De~Simoni, L. De~Lillo, M.~Ruzzene, and
  P.~Ermanni.
\newblock Phononic crystal with adaptive connectivity.
\newblock {\em Adv. Mater.}, 26(9):1343--1347, 2014.

\bibitem{Wen_JIMSS_2016}
J.~Wen, S.~Chen, G.~Wang, D.~Yu, and X.~Wen.
\newblock Directionality of wave propagation and attenuation in plates with
  resonant shunting arrays.
\newblock {\em J. Intell. Mater. Syst. Struct.}, 27(1):28--38, 2016.

\bibitem{Sugino_SMS_2017}
C.~Sugino, S.~Leadenham, M.~Ruzzene, and A.~Erturk.
\newblock An investigation of electroelastic bandgap formation in locally
  resonant piezoelectric metastructures.
\newblock {\em Smart Mater. Struct.}, 26(5):055029, 2017.

\bibitem{Collet_IEEESensors_2014}
M.~Collet, M.~Ouisse, and F.~Tateo.
\newblock Adaptive metacomposites for vibroacoustic control applications.
\newblock {\em IEEE Sens. J.}, 96(7):2145--2152, 2014.

\bibitem{suppl}
See Supplemental Material (in tail of this document) for a more detailed
  account on the experimental setups and for additional results. It includes
  Ref.~\cite{Viana_JBSMSE_2006, HUANG_IJES_2009}.

\bibitem{Wang_SMS_2010}
G.~Wang, S.~Chen, and J.~Wen.
\newblock Low-frequency locally resonant band gaps induced by arrays of
  resonant shunts with antoniou's circuit: experimental investigation on beams.
\newblock {\em Smart Mater. Struct.}, 20(1):015026, 2011.

\bibitem{dellISOLA_SMS_2004}
F.~dell'Isola, C.~Maurini, and M.~Porfiri.
\newblock Passive damping of beam vibrations through distributed electric
  networks and piezoelectric transducers: prototype design and experimental
  validation.
\newblock {\em Smart Mater. Struct.}, 13(2):299, 2004.

\bibitem{Tallarico_JMPS_2017}
D.~Tallarico, N.~V. Movchan, A.~B. Movchan, and D.~J. Colquitt.
\newblock Tilted resonators in a triangular elastic lattice: Chirality, bloch
  waves and negative refraction.
\newblock {\em J. Mech. Phys. Solids}, 103:236--256, 2017.

\bibitem{JIN_PRB_2017}
Y.~Jin, B.~Bonello, R.~P. Moiseyenko, Y.~Pennec, O.~Boyko, and
  B.~Djafari-Rouhani.
\newblock Pillar-type acoustic metasurface.
\newblock {\em Phys. Rev. B}, 96:104311, 2017.

\bibitem{FloresParra_SMS_2017}
E.~A. Flores~Parra, A.~Bergamini, L.~Kamm, P.~Zbinden, and P.~Ermanni.
\newblock Implementation of integrated 1d hybrid phononic crystal through
  miniaturized programmable virtual inductances.
\newblock {\em Smart Mater. Struct.}, 26(6):067001, 2017.

\bibitem{Viana_JBSMSE_2006}submission id: submit/2163362
F.~A.~C. Viana and V.~Steffen~Jr.
\newblock Multimodal vibration damping through piezoelectric patches and
  optimal resonant shunt circuits.
\newblock {\em {J. Braz. Soc. Mech. Sci. Eng.}}, 28:293--310, 2006.

\bibitem{HUANG_IJES_2009}
H.~H. Huang, C.~T. Sun, and G.~L. Huang.
\newblock On the negative effective mass density in acoustic metamaterials.
\newblock {\em Int. J. Eng. Sci.}, 47(4):610--617, 2009.

\end{thebibliography}
}
\end{document}